\documentclass[aps, pre, twocolumn, a4paper, floatfix]{revtex4-1}

\usepackage[utf8]{inputenc}
\usepackage{epsfig}
\usepackage[T1]{fontenc}
\usepackage{t1enc}
\usepackage{graphicx}
\usepackage{amssymb}
\usepackage{amsmath}

\usepackage{relsize}
\usepackage{color}
\usepackage{bm}
\usepackage[section]{placeins}
\usepackage{array}
\newcommand{\avg}[1]{{\left<#1\right>}}

\newcommand{\ceil}[1]{{\lceil #1\rceil}}

\begin{document}

\title{Emergence of robustness against noise: A structural phase
  transition in evolved models of gene regulatory networks}

\author{Tiago P. Peixoto}
\email{tiago@itp.uni-bremen.de}
\affiliation{Institut f\"{u}r Theoretische Physik, Universit\"at Bremen,
Otto-Hahn-Allee 1, D-28359 Bremen, Germany}


\begin{abstract}
  We investigate the evolution of Boolean networks subject to a
  selective pressure which favors robustness against noise, as a model
  of evolved genetic regulatory systems. By mapping the evolutionary
  process into a statistical ensemble and minimizing its associated free
  energy, we find the structural properties which emerge as the
  selective pressure is increased and identify a phase transition from a
  random topology to a ``segregated core'' structure, where a smaller
  and more densely connected subset of the nodes is responsible for most
  of the regulation in the network. This segregated structure is very
  similar qualitatively to what is found in gene regulatory networks,
  where only a much smaller subset of genes --- those responsible for
  transcription factors --- is responsible for global regulation. We
  obtain the full phase diagram of the evolutionary process as a
  function of selective pressure and the average number of inputs per
  node. We compare the theoretical predictions with Monte Carlo
  simulations of evolved networks and with empirical data for
  \textit{Saccharomyces
    cerevisiae} and \textit{Escherichia coli}.
\end{abstract}

\maketitle

\section{Introduction}

Many large-scale dynamical systems are composed of elementary units
which are \emph{noisy}, i.e., can behave non-deterministically, but
nevertheless must behave globally with some degree of predictability.  A
paradigmatic example is gene regulation in the cell, which is a system
of many interacting agents --- genes, mRNA, and proteins --- which are
subject to stochastic fluctuations. What makes gene regulation
particularly interesting is that it is assumed to be under evolutionary
pressure to preserve its dynamic memory against stochastic
fluctuations~\cite{kitano_biological_2004, raser_noise_2005,
  maheshri_living_2007}. Many important cellular processes require such
reliability, such as circadian
oscillations~\cite{raser_noise_2005}. Furthermore, in multicellular
organisms, errors in signal transduction can potentially lead to
catastrophic consequences, such as embryo defects or
cancer~\cite{kitano_biological_2004, gutkind_signaling_2000}. Since the
source of noise cannot be fully removed~\cite{lestas_fundamental_2010},
a gene regulation system must adopt characteristics which compensate for
the unavoidably noisy nature of its elements. Since they are a product
of natural selection, these characteristics must emerge from random
mutations and subsequent selection based on fitness. A central question
concerns the general large-scale features which are likely to emerge in
this scenario that result in reliable function under noise. In this
work, we study the emergence of robustness against noise in networks of
Boolean elements which are subject to selective pressure, functioning as
a model for evolved gene regulatory systems. We show that the system
undergoes a structural phase transition at a critical value of selective
pressure, from a totally random topology to a ``segregated-core''
structure, where a smaller and more densely connected subset of the
network is responsible for the regulation of most nodes in the
network. This characteristic is present to a significant degree in gene
regulatory systems of organisms such as yeast and \emph{Escherichia
coli}, in which all the regulation is done by a much smaller (and
denser) subset of the network, comprised of transcription factor genes.

Boolean networks (BNs) have been used extensively to model gene
regulation~\cite{kauffman_metabolic_1969, kauffman_homeostasis_1969,
  bornholdt_systems_2005, drossel_random_2008}. The Boolean value on a
given node represents the level of concentration of proteins encoded by
a gene, which in the simplest approximation can be either ``on'' or
``off.'' The regulation of genes by other genes is represented by
Boolean functions associated with each node, which depend on the state
of other nodes called the \emph{inputs} of the function. The dynamics on
these networks serve as a model for the mutual regulation of genes which
control the metabolism of cells in an organism. Gene regulation is
composed of specific steps involving the production of proteins and
other metabolites, which need to be carried out in specific sequences
and under certain conditions. During each of these steps the dynamics is
subject to stochastic fluctuations~\cite{raser_noise_2005,
  maheshri_living_2007, raj_nature_2008}, since the number of proteins
involved can be very low~\cite{raser_noise_2005, eldar_functional_2010},
and the whole process lacks an inherent synchronization mechanism. In
order for the regulation process to work reliably, the network must
possess some degree of robustness against these
perturbations~\cite{kollmann_design_2005}. Indeed, the investigation of
real regulatory networks modelled as BNs, such as the one responsible
for the yeast cell cycle~\cite{li_yeast_2004}, revealed a remarkable
degree of robustness, where most trajectories in state space lead to the
same attractor, regardless of the initial conditions. Similar results
were also obtained for the segment polarity regulatory network in
\emph{Drosophila melanogaster}~\cite{albert_topology_2003,
chaves_robustness_2005}, which showed that wild-type attractors are
significantly robust to different initial conditions and perturbations,
and seem to depend only on general topological characteristics of the
network, instead of specific functional details. However, the general
features which make BNs robust against different types of perturbations
are still being identified.

Perhaps the simplest form of perturbation one can consider is a ``flip''
of a single node in the network, and the propagation of flips which
result from it. This corresponds to the situation where the stochastic
noise is very weak, and can be modeled as a single flip event. After
the perturbation, the system has an arbitrary amount of time to recover
(if it recovers), and different perturbations do not build up. Many
authors have considered the robustness against perturbations of this type,
including Kauffmann~\cite{kauffman_metabolic_1969,
  kauffman_homeostasis_1969} who was the first to propose random Boolean
networks (RBNs) --- networks with fully random topology and functions
--- as a model of gene regulation. According to this type of
perturbation, the dynamics of RBNs~\cite{drossel_random_2008} can belong
to one of two phases, depending on the number of inputs per node $K$: A
frozen phase ($K<2$) where the perturbation propagates sub-linearly in
time and eventually dies out; and a ``chaotic'' phase ($K>2$) where the
perturbation grows exponentially and eventually reaches the entire
system. A critical line exists at $K=2$, where the perturbation grows
algebraically, and features from both phases are simultaneously
observed.

Although RBNs in the frozen phase and on the critical line show features
which can be interpreted as robustness in some sense, they fall short of
being convincing models for gene regulation. Actual gene regulation
networks are not random and show a high degree of
topological~\cite{maslov_computational_2005} and
functional~\cite{harris_model_2002} organization which are not present
in simple RBNs. Conceivably, these features arise out of stringent
requirements to perform specific tasks and of types of robustness which
are more demanding than the containment of single flip perturbations. As
an attempt at producing more complete models, many authors investigated
the evolution of BN systems, where the fitness criterion is some form of
robustness against perturbation which is not inherent to RBNs. The
majority of authors assumed single flips as the only type of noise, but
considered different types of responses as fitness
criteria~\cite{bornholdt_neutral_1998, bornholdt_robustness_2000,
  stern_emergence_1999, bassler_evolution_2004, aldana_robustness_2007,
  szejka_evolution_2007, mihaljev_evolution_2009,
  pomerance_effect_2009}, most of which are related to the capacity of
the network to display the same dynamical pattern after a
single-flip. In particular, in~\cite{szejka_evolution_2007} it was found
that if the fitness criterion is the ability to return to the same
attractor after the perturbation, the evolved networks always achieve
maximum fitness. Furthermore these networks with maximum fitness span a
huge portion of configuration space, and show a high degree of
variability. This means not only that this type of robustness can
evolve, but also that it is not a very demanding task for the
evolutionary process.

In this work, we consider the arguably more realistic situation where
the perturbations are caused by transcriptional noise, which can be
arbitrarily frequent~\footnote{Another realistic source of noise are
  perturbations in the update sequence of nodes, since gene regulation
  lacks a global synchronizing
  clock~\cite{klemm_topology_2005}. However, it can be shown that
  absolute robustness against this type of noise can be achieved in an
  independent manner, and with a very small effect to the global
  topological characteristics of the
  system~\cite{peixoto_boolean_2009}.}. In this scenario, the effects of
noise can overlap and build up in time. The appropriate fitness
criterion remains whether or not the network is capable of performing
some predefined dynamical pattern, but this is a task which becomes much
more complicated. In fact, it can be shown that for networks which are
sparse, i.e., the average number of inputs per node is some finite
number, perfect robustness can never be achieved, and some amount of
deviations, or ``errors,'' in the dynamics are always going to
exist~\cite{peixoto_redundancy_2010, peixoto_behavior_2012}. Instead,
one measures robustness not only by the amount of existing errors, but
also by the ability of the system to not be overtaken by them and
consequently lose all memory of its dynamical past --- i.e., to become
ergodic. This type of robustness is much stronger than, and not
necessarily related to the ability of the system to contain single-flip
perturbations. This was shown in~\cite{peixoto_noise_2009} for RBNs
subject to transcriptional noise, for which neither phase (chaotic or
frozen) is robust, and both display ergodic behavior, for any non-zero
value of noise.

Furthermore, unlike~\cite{bornholdt_neutral_1998,
  bornholdt_robustness_2000, stern_emergence_1999,
  bassler_evolution_2004, szejka_evolution_2007,
  mihaljev_evolution_2009}, in this work we also consider the
\emph{cost} which is associated with different levels of robustness. It
is generally the case that improved robustness can be obtained by
introducing redundancy or some other mechanism that counteracts the
effect of noise, which increases the overhead in the system. This added
overhead can impact negatively on the fitness of the organism, which
needs to spend more energy or more time to perform the same
task. Therefore the trade-off between overhead and robustness is also
driven by the evolutionary process. In this work, this overhead is
controlled by fixing the average in-degree during the evolutionary
process, which becomes an external parameter. By selecting the
appropriate value, one automatically determines a selective pressure
that yields the corresponding trade-off.

Our main result is that under transcriptional noise, the selective
pressure can have a very noticeable effect on large-scale properties of
the system: If it is large enough, it triggers a structural phase
transition, where networks change from a random topology to a
segregated-core structure, with most nodes being regulated by a smaller
and denser subset of the network. This observed segregated-core topology
is strikingly similar (even if qualitatively so) to what is observed in
most real gene regulation networks; namely, genes are separated into two
classes: target genes, and those which regulate transcription
factors. Only transcription factor genes are responsible for regulation
of other genes, and they are usually orders of magnitude smaller in
number than target genes~\cite{nimwegen_scaling_2006}.

This work is divided as follows. We begin in Sec.~\ref{sec:model} by
presenting the model, and in Sec.~\ref{sec:evolution} we define the
evolutionary process and map it into an equivalent Gibbs ensemble. We
then parametrize the topological characteristics of the system as a
stochastic blockmodel in Sec.~\ref{sec:block}, and obtain an expression
for its entropy. In Sec.~\ref{sec:free-energy} we describe the technique
used to minimize the free energy. We follow in Sec.~\ref{sec:phase} with
the characterization of the existing phase transition and obtain the
phase diagram. We perform comparisons with Monte Carlo simulations in
Sec.~\ref{sec:mc} and with the gene regulatory networks of yeast and
\emph{E. coli} in Sec.~\ref{sec:real}. Finally, we conclude with a
discussion.

\section{The model}\label{sec:model}

A Boolean network~\cite{kauffman_metabolic_1969,drossel_random_2008} is
a directed graph of $N$ nodes representing Boolean variables
$\mathbf{\sigma} \in \{1,0\}^N$, which are subject to a deterministic
update rule,
\begin{equation}\label{eq:bn_dyn}
  \sigma_i(t+1) = f_i\left(\bm{\sigma}(t)\right)
\end{equation}
where $f_i$ is the update function assigned to node $i$, which depends
exclusively on the states of its inputs. At a given time step all nodes
are updated in parallel.

We include noise in the model by introducing the probability $P$ that at
each time step a given input has its value ``flipped'': $\sigma_j \to 1
- \sigma_j$, before the output is
computed~\cite{peixoto_noise_2009}. This probability is independent for
all inputs in the network, and many values may be flipped
simultaneously. The functions on all nodes are taken to be the majority
function, defined as
\begin{equation}\label{eq:maj}
  f_i(\{\sigma_j\}) = 
  \begin{cases}
    1  \text{ if } \sum_j\sigma_j > k_i / 2, \\
    0  \text{ otherwise, }
  \end{cases}
\end{equation}
where $k_i$ is the number of inputs of node $i$. It is assumed
throughout the paper that the values of $k_i$ are always odd~\footnote{
  The definition above will lead to a bias if $k_i$ is an even number,
  since if the sum happens to be exactly $k_i / 2$ the output will be
  $0$, arbitrarily. Alternative definitions could be used, which would
  remove the bias~\cite{szejka_phase_2008}. However, for the analysis
  presented here, this is not an issue since $k_i$ is always odd.}.
This is so chosen because odd-valued majority functions are optimal,
since no other function performs better against
noise~\cite{evans_maximum_2003}. By using Eq.~\ref{eq:maj}, we
essentially remove the choice of functions from the evolutionary
process and concentrate solely on topological aspects.

Starting from an initial configuration, the dynamics of the system
evolves and eventually reaches a dynamically stable regime, where the
average fraction $b_t$ of nodes with value $1$ no longer changes, except
for stochastic fluctuations which vanish for a large system
size~\cite{huepe_dynamical_2002, peixoto_behavior_2012}. In the absence
of noise ($P=0$) there are only two possible attractors (if the network
is sufficiently well connected) where all nodes have the same value,
which can be either $0$ or $1$. We will consider these homogeneous
attractors as representing the ``correct'' dynamics, and denote the
deviations from them as ``errors.'' More specifically, and without loss
of generality, we will name the value of $1$ as an ``error,'' and the
value of $b_t$ as the average error on the system.

The steady-state fraction of errors $b^* \equiv \lim_{t\to\infty} b_t$
(for $b_0=0$) will increase with $P$. For any network with a finite
average in-degree there will be a critical value of noise $P^*$ for
which the dynamics undergoes a phase transition, and the value of $b^*$
reaches $1/2$, and remains at this value for
$P>P^*$~\cite{peixoto_redundancy_2010}. The value $b^*=1/2$ is special,
since it means that the dynamics lost the memory of its initial state,
since any other initial value of $b_0$ (including $b_0>1/2$) would lead
eventually to this same value of $b^*$. Therefore, the value of $P^*$
marks the transition from a nonergodic to an ergodic
dynamics. Robustness against noise is synonymous with nonergodicity,
since only in this regime are dynamical correlations not destroyed over
time.

BNs with majority functions serve as a paradigmatic model for networks
robust against noise, since they are composed of optimal elements, and
they show a minimal dynamical behavior in the absence of noise, namely
two homogeneous attractors with $\{\sigma_i\} = 0$ or $1$. If robustness
cannot be attained for such a system, it is much less likely to be
possible for a different system with a another choice of Boolean
functions, or displaying a more elaborate dynamical
pattern~\cite{peixoto_behavior_2012}.

In this work we will consider the value of the steady-state average
error $b^*$ as the main fitness criterion governing the survival
probability of an organism, since it directly measures the deviation
from the situation without noise. Although the phenotype itself,
i.e. the dynamics without noise, does not change during the evolutionary
process considered here, its \emph{stability}, as measured by $b^*$,
does. This translates into an actual fitness criterion, since it is not
enough for phenotypes to exist; they must also be stable against
perturbations. If they are not, they are not viable in practice, and
thus should not be observed.

\section{Evolutionary dynamics}\label{sec:evolution}

We suppose that a given BN represents the genotype of a full organism,
which can self-replicate and belongs to a population that is subject to
an evolutionary pressure. The number of individuals in the population is
assumed to be sufficiently large and constant. Individuals replicate a
given number of times with a constant rate. Parents die the moment they
replicate. The offspring are always initially identical to their
parents, but are individually subject to point mutations represented by
the matrix $\mu_{ij}$, which defines the probability of mutating from
genotype (i.e. network) $i$ to $j$. The offspring survive with
probability $a_i$, given by the Boltzmann selection criterion
\begin{equation}
  a_i \propto e^{\beta f_i},
\end{equation}
where $f_i$ is the fitness of genotype $i$. The parameter $\beta$
controls the \emph{selective pressure}: For large values of $\beta$ only
the very best networks survive, whereas for smaller values most networks
do. As mentioned previously, the fitness of a network will be given by
the fraction of ones (``errors'') after a sufficiently long time,
$b^*_{(i)} \equiv \lim_{t\to\infty} b_t^{(i)}$, for $b_0^{(i)}=0$, as
\begin{equation}
  f_i = -Nb^*_{(i)}.
\end{equation}
Thus, the largest fitness a network can have is $f_i=0$, which should
be possible only if there is no noise ($P=0$).

We suppose that the global offspring mortality rate is such that the
size of the population always remains constant. If we consider that the
dynamics occurs in discrete time steps, we can write the probability
$\pi_i(t)$ of finding an individual in the population with genotype $i$
at time $t$ as a Markov chain,
\begin{equation}\label{eq:markov}
  \pi_i(t) =  a_i \sum_j \pi_j(t-1) \mu_{ji}.
\end{equation}
The mutation probabilities $\mu_{ji}$ have a decisive effect on what
topologies emerge.  Mutations in actual biological systems may result in
topological bias, such as gene duplications, which are not reversible
and result in networks with broad degree
distributions~\cite{ispolatov_duplication-divergence_2005,
enemark_gene_2007}.  However, the central aim here it to obtain the most
likely topology that arises due to the selective pressure
\emph{alone}. For this reason we are more interested in mutations which
will lead to all possible networks with equal probability in the absence
of selective pressure (i.e. ergodicity). A simple and conventional
choice is reversible mutations, $\mu_{ij} = \mu_{ji}$, for which the
steady state $\pi_i \equiv \lim_{t\to\infty} \pi_i(t)$ obeys the
detailed balance condition: $\pi_i \mu_{ij} a_j = \pi_j \mu_{ji}
a_i$. This is a sufficient condition for the desired ergodicity
property, but it is not strictly necessary, since other types of
mutations may also fulfill it. However, from this condition we easily
obtain that the steady-state probability of finding an individual with
genotype $i$ is given by its survival probability,
\begin{equation}
  \pi_i = a_i =  e^{\beta f_i} / \mathcal{Z},
\end{equation}
where $\mathcal{Z} = \sum_i e^{\beta f_i} = \sum_i e^{-\beta
Nb^*_{(i)}}$.  This corresponds exactly to a Gibbs ensemble of all
possible genotypes, with a partition function given by $\mathcal{Z}$,
where $Nb^*_{(i)}$ plays the role of the ``microstate energy'' and
$\beta$ is the ``inverse temperature'' (these are of course only
mathematical analogies, since these quantities do not actually represent
a physical energy and temperature, respectively). The average intensive
``energy'' in the ensemble is thus
\begin{equation}\label{eq:bcan}
  b^* = \sum_i b^*_{(i)} e^{-\beta Nb^*_{(i)}}/\mathcal{Z},
\end{equation}
and the canonical entropy is
\begin{equation}\label{eq:scan}
  \mathcal{S} = -\sum_i\pi_i\ln\pi_i = \ln\mathcal{Z} + \beta Nb^*.
\end{equation}
The objective is to obtain not only $b^*$ for a given $\beta$, but also
the network topologies which characterize the ensemble. Instead of
considering all microstates individually (i.e., all possible networks)
and computing Eqs.~\ref{eq:bcan} and~\ref{eq:scan} directly, we may
parametrize the whole ensemble via some macroscopic variables $\{x_j\}$
which sufficiently describe its topological properties. These variables
must be chosen so that it is possible to write both $b^*(\{x_j\})$ and
$\mathcal{S}(\{x_j\})$ as functions of these variables alone. The
entropy can, for instance, be obtained via the microcanonical
formulation
\begin{equation}
  \mathcal{S}(\{x_j\})=\ln\Omega(\{x_j\}),
\end{equation}
where $\Omega(\{x_j\})$ is the number of different networks given a
macroscopic parametrization $\{x_j\}$.  The values of $\{x_j\}$ which
correspond to thermodynamic equilibrium [i.e., the steady state of
Eq.~\ref{eq:markov}] can be obtained by minimizing the ``free energy'',
\begin{equation}\label{eq:fe}
  \mathcal{F}(\{x_j\}) = Nb^*(\{x_j\}) - \mathcal{S}(\{x_j\})/\beta,
\end{equation}
with respect to $\{x_j\}$. This stems from the principles of maximum
entropy and minimum energy, for closed systems with fixed energy and
entropy, respectively, which need to hold in thermodynamic
equilibrium~\cite{callen_thermodynamics_1985}. It should again be
emphasized that the theory so far is only a mathematical tool, which,
although exact, says nothing about the actual physical thermodynamical
properties of the evolved systems, i.e., they have no relation to an
actual measurable energy or temperature. Instead, the minimization of
Eq.~\ref{eq:fe} is entirely analogous to obtaining the steady state of
Eq.~\ref{eq:markov} by any other means.  However, this approach,
together with an appropriate topological parametrization, allows us to
obtain the outcome of the evolutionary process on the population,
without having to actually implement any dynamics.  As will be described
in detail in the next section, we will parametrize the ensemble as a
general \emph{stochastic blockmodel}, which allows for a wide range of
topological configurations, while at the same time allowing for a
tractable computation of $b^*$ and $S$, which then can be used to
minimize $\mathcal{F}$.

It should also be mentioned at this point that we are interested in the
properties of typical networks in the ensemble when the selective
pressure $\beta$ is varied, under the restriction that the average
number of inputs per node (the average in-degree) $\avg{k}$ is always
the same. As mentioned in the introduction, this restriction originates
from the assumption that a larger number of inputs increases the
putative cost for the organism of realizing a regulatory mechanism which
depends on more elements. Thus, the value of $\avg{k}$ should on its own
impact the fitness of the organism, and should also be subject to
natural selection. For simplicity, we do not describe the fitness
landscape which depends on $\avg{k}$ and its evolution, in order to
emphasize the effects of robustness against noise alone. Instead, we
consider $\avg{k}$ as an external parameter, which essentially means
that the fitness pressure on $\avg{k}$ supersedes that of the other
parameters, such that it cannot change during evolution. In this way, we
are implicitly considering the cost associated with the robustness
achieved by increasing $\avg{k}$: the smaller is the value of $\avg{k}$
chosen, the larger is the implied fitness penalty of having more
connections.

\section{Stochastic blockmodel}\label{sec:block}

Simultaneous consideration of all possible networks with a given
$\avg{k}$ is a tremendous task, due to the gigantic number of diverse
configurations which are possible. For arbitrary networks the
computation of $b^*$ according to Eq.~\ref{eq:bn_dyn} may be very
cumbersome, since it may depend on many degrees of freedom. Therefore,
we narrow down the allowed subset of possible network topologies to
those which can be accommodated in a \emph{stochastic
  blockmodel}~\cite{holland_stochastic_1983,faust_blockmodels:_1992,karrer_stochastic_2011}
\footnote{Blockmodels are essentially equivalent to the hidden-variable
  model~\cite{boguna_class_2003}, when the hidden variables are
  discrete, and their multiplicity is smaller than the number of
  nodes.}. As will become clear in the following, we do so without
sacrificing the generality of the approach, since we can progressively
add to this model as many degrees of freedom as we desire, and in this
way obtain arbitrarily elaborate structures in a controlled fashion.

\begin{figure} [tb]
  \vspace{-3em}
  \begin{center}
    \includegraphics*[width=0.5\columnwidth]{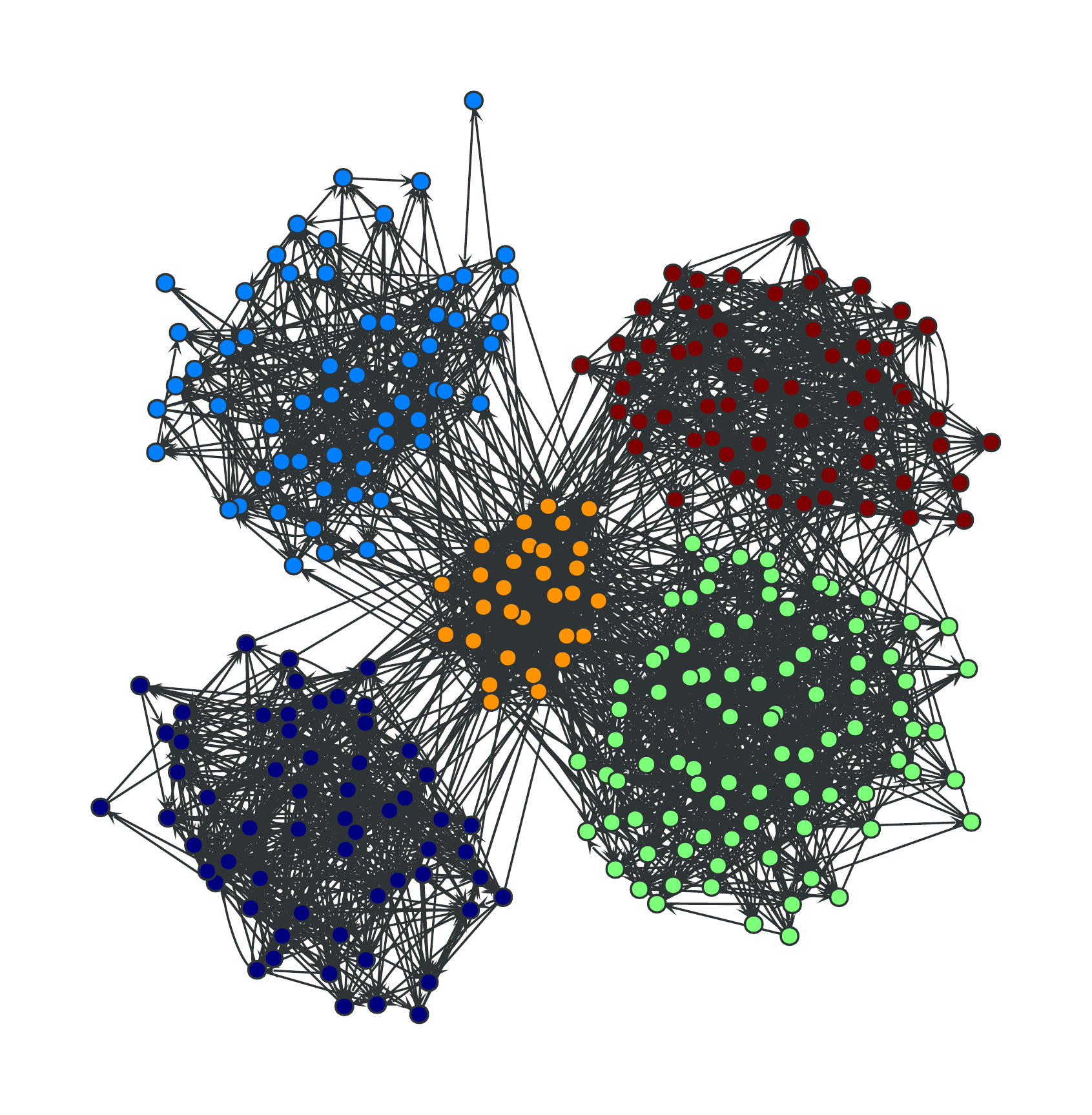}
  \end{center} \caption{\label{fig:blocks}(Color online) Example of a
  network corresponding to a blockmodel with five blocks. The vertices
  of each block are labeled with the same color.}
\end{figure}

A stochastic blockmodel assumes that the nodes in the network can be
partitioned into discrete \emph{blocks}, such that every node belonging
to the same block has (on average) the same characteristics. Hence, for
very large systems, we need only to describe the degrees of freedom
associated with the individual blocks (see Fig.~\ref{fig:blocks}). By
considering a system composed of many blocks, we can describe a wide
array of possible topological configurations.

More precisely, a (degree-corrected~\cite{karrer_stochastic_2011})
stochastic blockmodel is a system of $n$ blocks, where $w_i$ is the
fraction of nodes in the network which belong to block $i$ (we have
therefore that $\sum_iw_i=1$), and $p^i_k$ is the in-degree distribution
of block $i$. Thus, the average in-degree is given by $\avg{k} =
\sum_{k,i}kp^i_kw_i$. The matrix $w_{j\to i}$ describes the fraction of
the inputs of block $i$ which belong to block $j$ (we have therefore
that $\sum_jw_{j\to i}=1$). Since the out-degrees are not explicitly
required to describe the dynamics (see Eq.~\ref{eq:dynblock} below),
they will be assumed to be randomly distributed, subject only to the
restrictions imposed by $w_i$ and $w_{j\to i}$.

We note that, although we have diminished the class of networks which
will be accessible by the evolutionary algorithm, we still allow a very
large array of possible configurations, which can in principle
incorporate arbitrary in-degree distributions, degree
correlations~\cite{newman_structure_2003}, assortative or disassortative
mixing~\cite{newman_mixing_2003}, and community
structure~\cite{girvan_community_2002}, to name only a few
properties. As will become clear in the following section, this
blockmodel is sufficient to characterize the most important topological
property that is relevant for robustness against noise, which is the
formation of densely connected central subgraphs.

\subsection{The value of $b^*$ for a blockmodel}

Supposing that the number of vertices $Nw_i$ belonging to each block $i$
is arbitrarily large, we can compute the value of $b^*$ using an
heterogeneous version of the annealed
approximation~\cite{derrida_random_1986}, by supposing that at each time
step the inputs of each function are randomly chosen, such that the
specified block structure given by $w_{i\to j}$ is always
preserved~\cite{peixoto_behavior_2012}. If the number of vertices in
each block is large enough, we can expect this approximation to become
an exact description for quenched networks as well. We can then write
the average value of $b_i$ for each block over time as
\begin{equation}\label{eq:dynblock}
  b_i(t+1) = \sum_kp_k^im_k\left((1-2P)\sum_j w_{j\to i}b_j(t) + P\right),
\end{equation}
which is a system of $n$ coupled maps, where $m_k(b)$ is the probability
that the output of a majority function will be $1$, if the inputs are
$1$ with probability $b$, and is given by
\begin{equation}\label{eq:maj_p}
  m_k(b) = \sum_{i = \ceil{k/2}}^k{k \choose i} b^i(1-b)^{k-i}.
\end{equation}
A fixed point of Eq.~\ref{eq:dynblock} represents the solution of a
polynomial system of arbitrary order, and therefore cannot be written in
closed form. However, it can be obtained numerically by starting the
system at $b_i = 0$, and iterating Eq.~\ref{eq:dynblock} until a
fixed-point $\{b^*_i\}$ is reached. The value of $b^*$ can then be
obtained as $b^* = \sum_iw_ib^*_i$.

\subsection{Blockmodel entropy}\label{sec:entropy}
We obtain the entropy of the stochastic blockmodel
ensemble~\cite{bianconi_entropy_2009, peixoto_entropy_2011} by
enumerating all possible networks which are compatible with a given
choice of $w_i$, $p_k^i$ and $w_{i\to j}$. To make the counting simpler,
we ignore the difficulty of forbidding parallel edges, and consider the
ensemble of \emph{configurations}, since the occurrence of parallel
edges should vanish for large network sizes
(see~\cite{peixoto_entropy_2011} for more details).  Later we compare
the results obtained with Monte-Carlo simulations with parallel edges
forbidden, and we find very good agreement.

We begin by enumerating all possible in-degree sequences of each block
which correspond to the prescribed in-degree distributions,
\begin{equation}
  \Omega_d =\prod_i\frac{(Nw_i)!}{\prod_k{(Nw_ip_k^i)!}}.
\end{equation}
For a given block $i$ with a fixed in-degree sequence, we can count the
number of different input choices as
\begin{equation}
  \Omega^i_e = \frac{E_i!}{\prod_jE_{j\to i}!}\prod_j(Nw_j)^{E_{j\to i}},
\end{equation}
where $E_i=Nw_ik_i$ is the total number of inputs belonging to block $i$
and $E_{j\to i}=w_{j\to i}E_i$ is the total number of inputs from block
$i$ which belong to block $j$. Since the set of inputs of each function
is unordered, we still need to divide the whole number of input
combinations by $\prod_k(k!)^{N_k}$, where $N_k=N\sum_{i,k}w_ip_k^i$ is
the total number of vertices with in-degree $k$. Putting it all together
we have
\begin{equation}
    \Omega = \Omega_d \frac{\prod_i \Omega^i_e}{\prod_k(k!)^{N_k}}.
\end{equation}
Taking the logarithm of this expression, and the limit $N \gg 1$, and
using Stirling's approximation, we obtain the full entropy (up to a
trivial constant term, which is not relevant to the minimization of the
free energy),
\begin{multline}
  \mathcal{S}/N = \avg{k}\ln N + \sum_iw_i\mathcal{S}_i^k \\
  - \sum_iw_ik_i\sum_jw_{j\to i}\ln \left(\frac{w_{j\to i}}{w_j}\right),
\end{multline}
where $\mathcal{S}_i^k$ is an entropy term associated with the degree
distribution of block $i$, and is given by
\begin{equation}
  \mathcal{S}_i^k = -\sum_kp_k^i(\ln p_k^i + \ln k!).
\end{equation}

\subsection{Choice of single-block in-degree distribution}

We want to constrain the number of degrees of freedom in the model, such
that only the average in-degree $k_i$ of each block is specified, not
the entire distribution. In this way, graphs with many different global
in-degree distributions are still possible by composing different blocks
with different $k_i$'s, but we have a finite number of degrees of
freedom per block. In order to obtain the in-degree distribution of the
individual blocks, we maximize the entropy $\mathcal{S}$, with the
restriction that the average in-degrees are fixed. For that, we
construct the Lagrangian,
\begin{equation}
  \Lambda = \mathcal{S} - \sum_i\lambda_i'\left(\sum_kkp_k^i - k_i\right) - \sum_i\mu_i\left(\sum_kp^i_k-1\right),
\end{equation}
where $\{\lambda_i\}$ and $\{\mu_i\}$ are Lagrange multipliers which
keep the averages and the normalizations constant. We note that the sum
over $k$ is made only over \emph{odd} values of $k$, due to the imposed
restrictions on the majority function. Obtaining the critical point
$(\{\frac{\partial\Lambda}{\partial p^i_k}\},
\{\frac{\partial\Lambda}{\partial \lambda'_i}\},
\{\frac{\partial\Lambda}{\partial \mu_i}\}) = 0$, and solving for
$\{p^i_k\}$ one obtains,
\begin{equation}\label{eq:pk}
  p_k^i = \frac{1}{\sinh \lambda_i} \frac{\lambda_i^k}{k!},
\end{equation}
where $k_i = \lambda_i / \tanh \lambda_i$. Equation~\ref{eq:pk} is a
Poisson distribution, which is defined and normalized only for odd
values of $k$.

This choice of $p^i_k$ is not necessarily the optimal one. In fact, it
is possible to show that single-valued distributions with zero variance
tend to provide the best error
resilience~\cite{peixoto_behavior_2012}. Nevertheless, the improvement
over a Poisson distribution is \emph{very} small, and the definition of
Eq.~\ref{eq:pk} allows for the average $k_i$ to be continuously varied,
which is very practical for the optimization of the free energy.

\subsection{Block splitting, decrease of entropy and the necessary
  number of blocks}\label{sec:split}

For the blockmodel defined in this section, there are $2n + n^2$
variables which define the topology, where $n$ is the number of
blocks. In order for arbitrary topologies to be faithfully represented
by the model, one would need to make $n\to\infty$, which would render
this approach impractical. However, we will show that for the purpose at
hand, only a minimal number of \emph{two} blocks is sufficient to fully
characterize the evolutionary process, without relying on any
approximations. This is due the following two facts: 1. Any possible
value of $b^*$ can be obtained with only two blocks; 2. Any other
topology with the same $b^*$ will invariably have a lower entropy, and
thus a larger free energy. Thus the minimum of the free energy will
always lie on a two-block structure.

The first fact can be shown by construction: Consider a system of two
blocks, where one of them (the ``core'') is smaller and much denser, and
the remaining block has an average in-degree close to the global
average. The inputs of the core block belong mostly to the core itself,
as well as the inputs of the remaining block. By changing the density of
the core block, as well as the degree of input segregation, it can be
shown~\cite{peixoto_behavior_2012} that any possible value of $b^*$ can
be achieved~\footnote{This is not the only two-block structure which can
  generate arbitrary values of $b^*$. In~\cite{peixoto_behavior_2012} it
  is shown how a bipartite ``restoration'' structure also achieves this,
  albeit less efficiently.}.

The second fact can be shown by considering a system of many blocks, and
selecting any two blocks, $l$ and $m$. If all other blocks are kept
intact, it can be shown that the entropy will always be larger if these
two blocks are merged into an effective single block. This can be shown
by partially maximizing the entropy $S$ via the Lagrangian,
\begin{equation}
  \Lambda = \mathcal{S} - \mu\left(\sum_ik_iw_i - \avg{k}\right) - \sum_i\gamma_i\left(\sum_jw_{j\to i}-1\right),
\end{equation}
where $\mu$ and $\{\gamma_i\}$ are Lagrange multipliers which keep the
average in-degree and the normalization of $w_{j\to i}$ fixed,
respectively. Obtaining the critical point
$(\frac{\partial\Lambda}{\partial w_{l\backslash m}},
\frac{\partial\Lambda}{\partial k_{l\backslash m}},
\{\frac{\partial\Lambda}{\partial w_{{l\backslash m}\to j}}\},
\{\frac{\partial\Lambda}{\partial w_{j\to {l\backslash m}}}\}) = 0$, and
solving for $w_{l\backslash m}, k_{l\backslash m}, \{w_{j\to l\backslash
  m}\}, \{w_{l\backslash m\to j}\}$ one obtains,
\begin{gather}
  k_l = k_m \\
  \frac{w_{l\to j}}{w_l} =  \frac{w_{m\to j}}{w_m} \\
  w_{j\to l} = w_{j\to m}.
\end{gather}
This corresponds to the situation where the nodes from blocks $l$ and
$m$ are topologically indistinguishable, i.e. the outgoing and incoming
edges are randomly distributed among the nodes of both blocks. Since any
arbitrary many-block structure can be converted into a single-block by
successive block merges, it follows directly that any arbitrary
many-block structure can be constructed by starting from a single block,
and successively splitting blocks. Thus, since the merging of blocks
always increases entropy, and the splitting decreases it, the entropy of
the final structure must be smaller than that of both the initial single
block and the succeeding two-block network.

In order for a many-block structure to have a lower free energy than the
two-block structure with the same value of $b^*$, it needs to have a
larger entropy. But according to the above argument, networks with a
larger number of blocks tend to have \emph{smaller} entropy. Networks
with larger entropy and number of blocks would have to be more
randomized than the two-block structure, which would invariably result
in a larger value of $b^*$. We can therefore conclude that the global
minimum of the free energy always occurs for a two-block structure, and
thus we need to deal with only eight variables~\footnote{We have empirically
  verified this by minimizing the free energy with up to 10 blocks, and
  the solutions were always \emph{identical} to that of the two-block
  case presented in the following section. The general character of the
  two-block topology was also verified by Monte Carlo simulations with
  up to $20$ blocks, as discussed later in the text (see also
  Fig.~\ref{fig:mc-multiblock}).}.

\section{Minimization of the free energy}\label{sec:free-energy}

Although we have an analytical expression for the entropy $\mathcal{S}$,
the value of $b^*$ cannot be obtained in closed form, and thus the same
is true for $\mathcal{F}$. Therefore we must resort to a numerical
computation of $b^*$, via the iteration of Eq.~\ref{eq:dynblock}, and
minimize $\mathcal{F}$ with a gradient descent algorithm, using finite
differences. Many of these methods work only for unconstrained
optimization problems, and we need to impose several constraints: The
average in-degree must be fixed, and the $w_i$ and $w_{j\to i}$
distributions must be normalized. However we can make the problem
unconstrained by using the following transformations,
\begin{gather}
  w_i = \frac{e^{\widetilde{w}_i}}{\sum_j e^{\widetilde{w}_j}} \\
  w_{j\to i} = \frac{e^{\widetilde{w}_{j\to i}}}{\sum_l e^{\widetilde{w}_{l\to i}}},
\end{gather}
where $\widetilde{w}_i$ and $\widetilde{w}_{j\to i}$ are unconstrained
real variables. Likewise we can transform $\lambda_i$ as
\begin{gather}
  \widetilde{k}_i = \frac{e^{\widetilde{\lambda}_i}}{\tanh e^{\widetilde{\lambda}_i}} \\
  k_i = c \widetilde{k}_i + \gamma \\
  \lambda_i = k_i \tanh \lambda_i  \label{eq:li},
\end{gather}
where $\widetilde{\lambda}_i$ is also an unconstrained real variable.
To obtain $\lambda_i$ from Eq.~\ref{eq:li} it is simply iterated until
it converges to the correct value, within some desired precision. The
values of $c$ and $\gamma$ are chosen to force $k_i \ge 1$ and the
average in-degree to a prescribed value $\avg{k}$,
\begin{equation}
  \begin{aligned}
  c &= \frac{\avg{k}}{\sum_i\widetilde{k}_i w_i},& \gamma &= 0,  &&\text{ if } \widetilde{k}_m \ge 1 \\
  c &= \frac{\avg{k} - 1}{\sum_i\widetilde{k}_i w_i - \widetilde{k}_m},& \gamma &= 1 - c\widetilde{k}_m,  &&\text{ otherwise, }  
\end{aligned}
\end{equation}
where $\widetilde{k}_m = \min(\{\widetilde{k}_i\})$. Thus we have
obtained an unconstrained minimization problem of function $\mathcal{F}$
with respect to the variables $\{\widetilde{w}_i\}$,
$\{\widetilde{w}_{j\to i}\}$, $\{\widetilde{\lambda}_i\}$.

In order to find the minimum of Eq.~\ref{eq:fe}, we employed the L-BFGS
quasi-Newton algorithm~\cite{byrd_limited_1995}, with the gradient
computed by finite differences.

\section{Structural phase transition}\label{sec:phase}

\begin{figure} 
  \vspace{-2em}
  \begin{center}
    \begin{minipage}{0.33\columnwidth}
      \begin{center}
        \includegraphics*[width=\columnwidth]{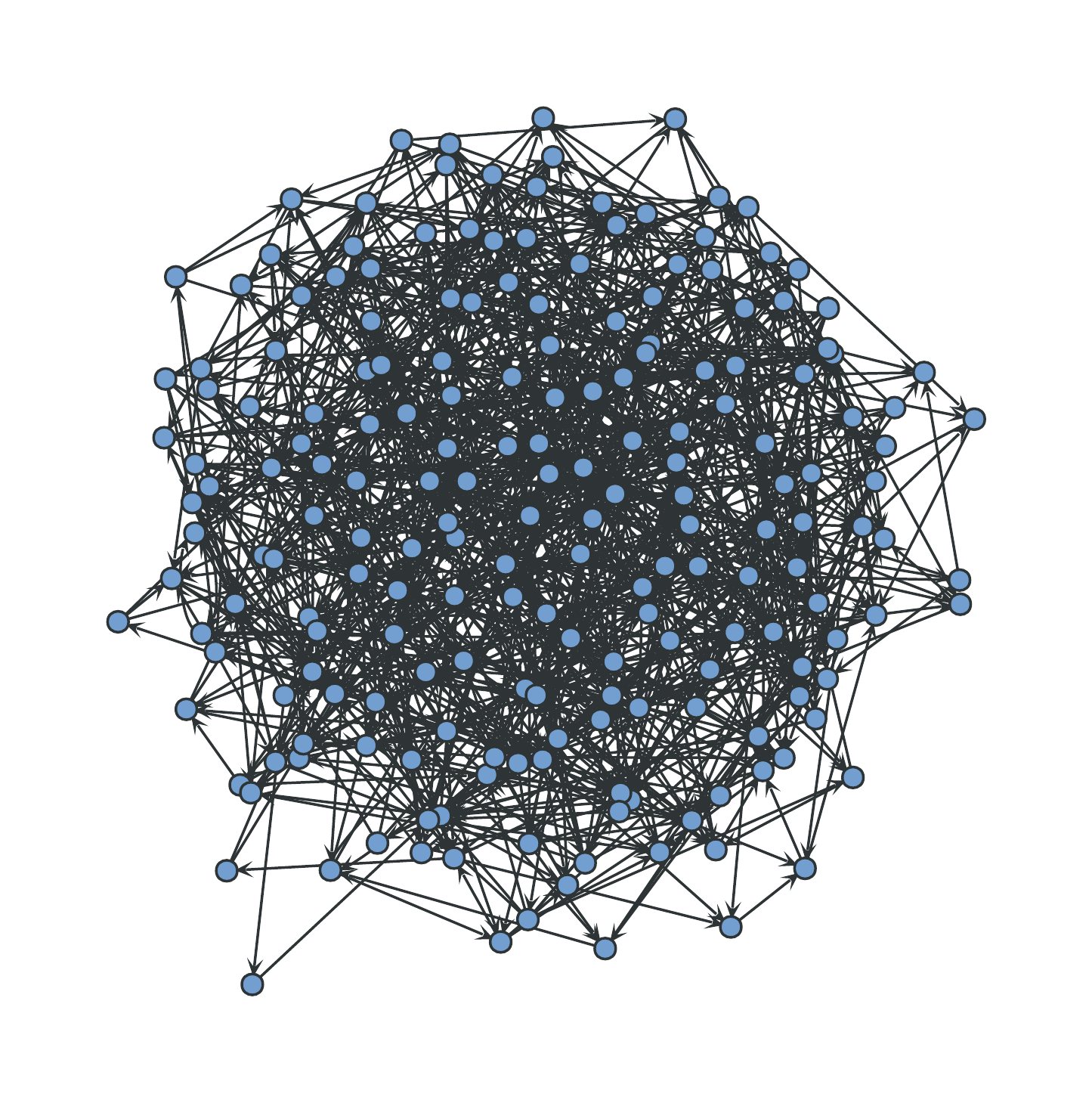}\\
        Random topology
      \end{center}
    \end{minipage}
    $\xrightarrow{\text{increasing } \beta}$
    \begin{minipage}{0.33\columnwidth}
      \begin{center}
        \includegraphics*[width=\columnwidth]{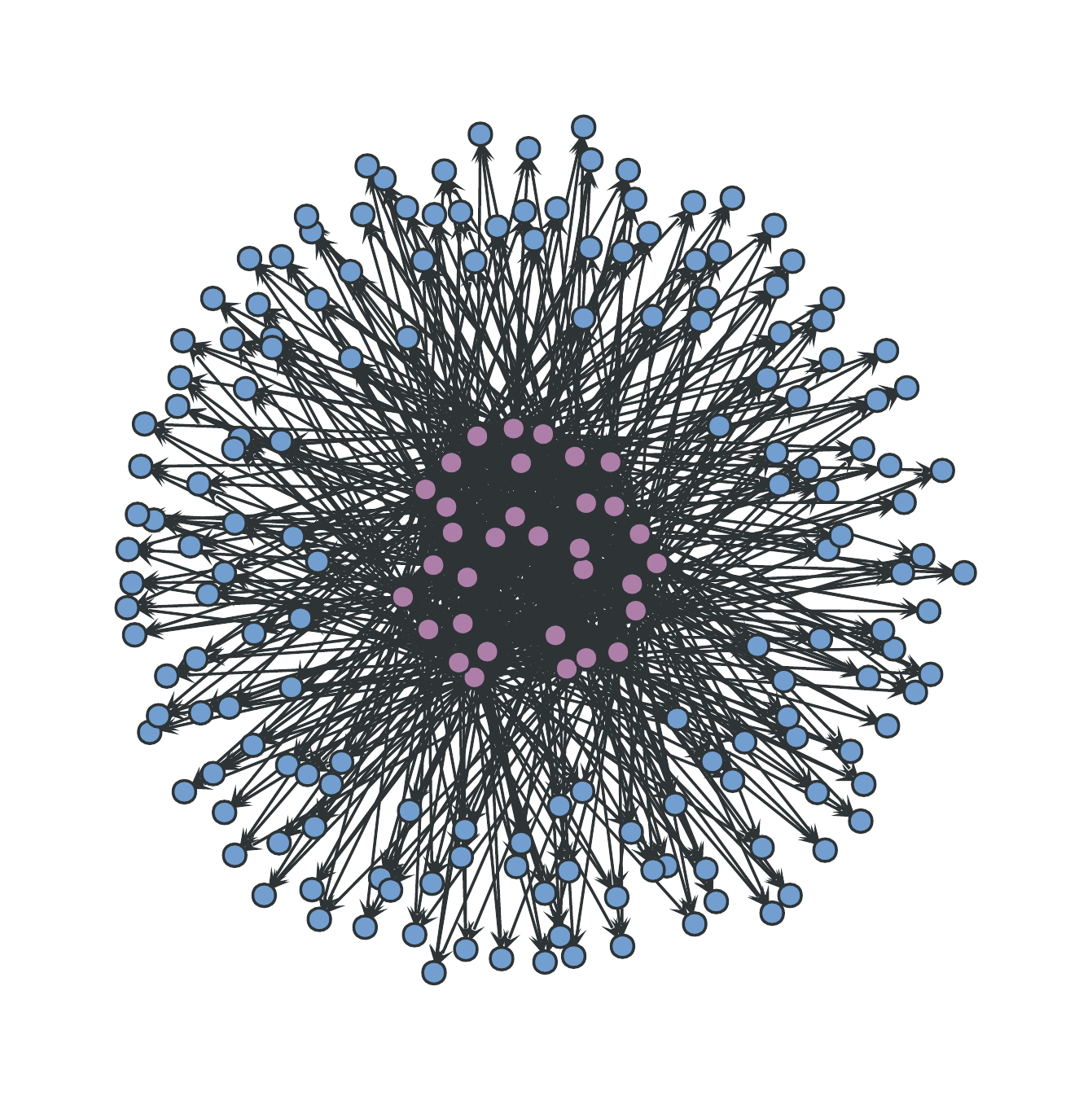}\\
        Segregated core
      \end{center}
    \end{minipage}
  \end{center}
  \caption{\label{fig:trans}(Color online) Structural phase transition observed when
    varying the selective pressure $\beta$, as described in the text.}
\end{figure}

The minimization of the free energy leads to one of two possible
structures (see Fig.~\ref{fig:trans}): 1. For low values of $\beta$ the
topology is a fully random graph; 2. For larger values of $\beta$ there
is the emergence of a \emph{segregated core} structure, where one of the
blocks has a larger in-degree density and is more responsible for the
regulation of the whole network.

In order to precisely characterize the phase transition, we define the
following order parameter,
\begin{equation}
  \phi = \frac{b^* - b_r}{b_{\text{min}} - b_r},
\end{equation}
where $b_r$ is the value of $b^*$ for a fully random network, and
$b_{\text{min}}$ is the smallest possible value of $b^*$ for a given
$\avg{k}$~\cite{peixoto_behavior_2012}, given by
\begin{equation}
  b_{\text{min}} = \sum_k p_k m_k(P),
\end{equation}
where $p_k$ is given by Eq.~\ref{eq:pk} with $k_i = \avg{k}$. We have
therefore that $\phi \in [0, 1]$ and if $\phi = 0$ the network ensemble
must be fully random, and if $\phi = 1$ it has the largest possible
value of fitness.

\begin{figure} 
  \flushleft
  \includegraphics*[width=0.49\columnwidth]{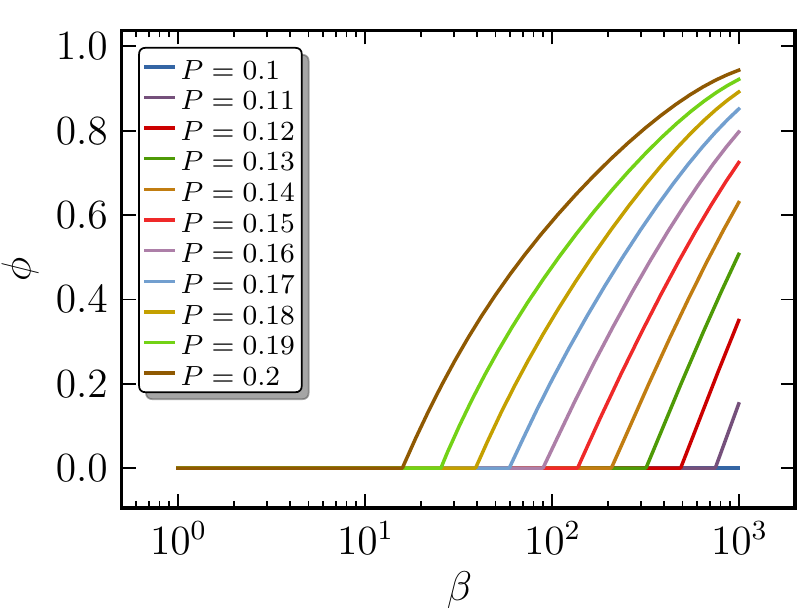}
  \includegraphics*[width=0.49\columnwidth]{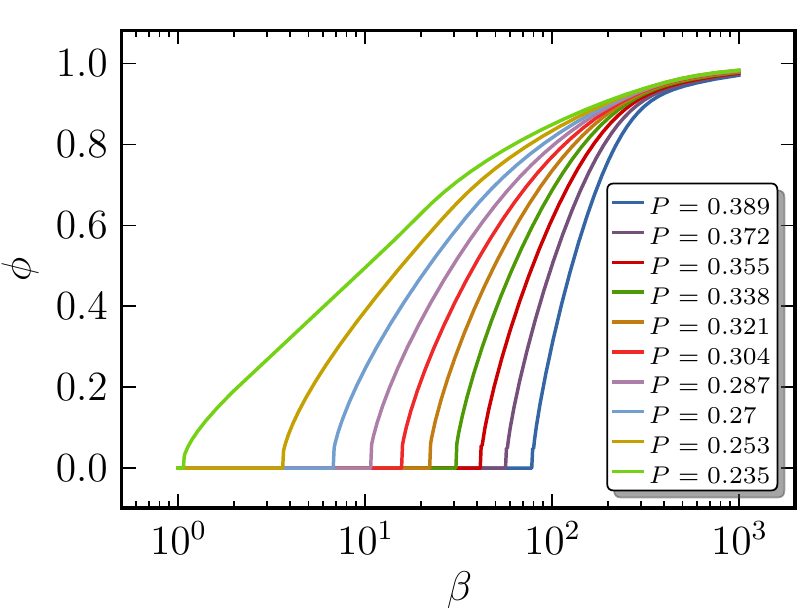}
  \caption{\label{fig:diag-op}
    (Color online) The order parameter $\phi$ as a function of the
    selective pressure $\beta$, for different noise levels $P$, and for
    $\avg{k} = 5$. The left panel shows curves for $P < P^*_r$, where
    $P^*_r$ is the critical value of noise for a fully random network,
    and the right panel shows curves for $P > P^*_r$. The curves on the
    left panel are shown in order of increasing $P$ from right to
    left, and on the right panel, from left to right.}
\end{figure}

\begin{figure} 
  \flushleft
  \begin{minipage}{0.49\columnwidth}
    \begin{center}
      \includegraphics*[width=1\columnwidth]{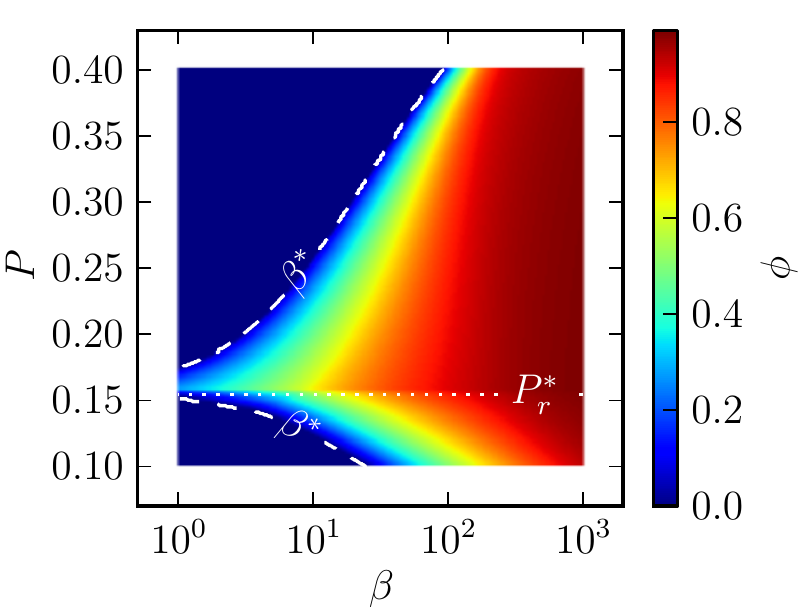}\\
      {\smaller[2] $\avg{k} = 3$}
    \end{center}
  \end{minipage}
  \begin{minipage}{0.49\columnwidth}
    \begin{center}
      \includegraphics*[width=1\columnwidth]{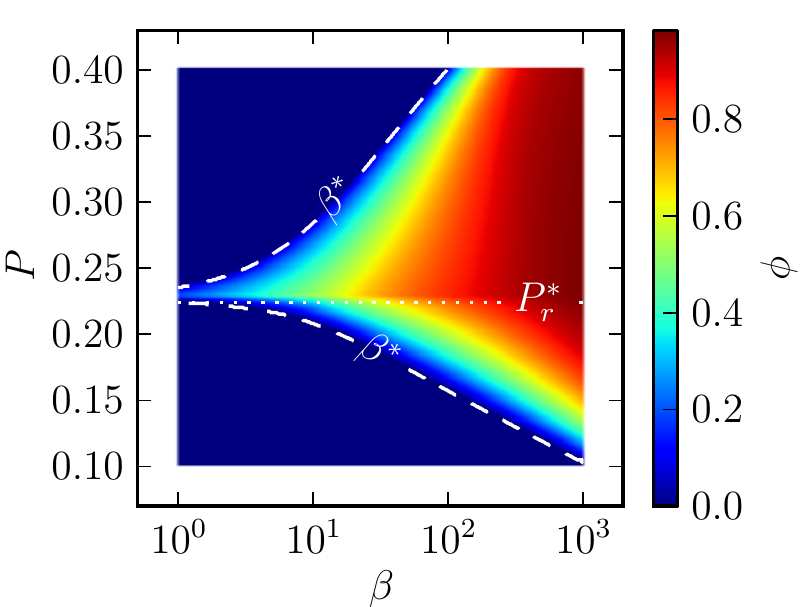}\\
      {\smaller[2] $\avg{k} = 5$}
    \end{center}
  \end{minipage}\\
  \begin{minipage}{0.49\columnwidth}
    \begin{center}
      \includegraphics*[width=1\columnwidth]{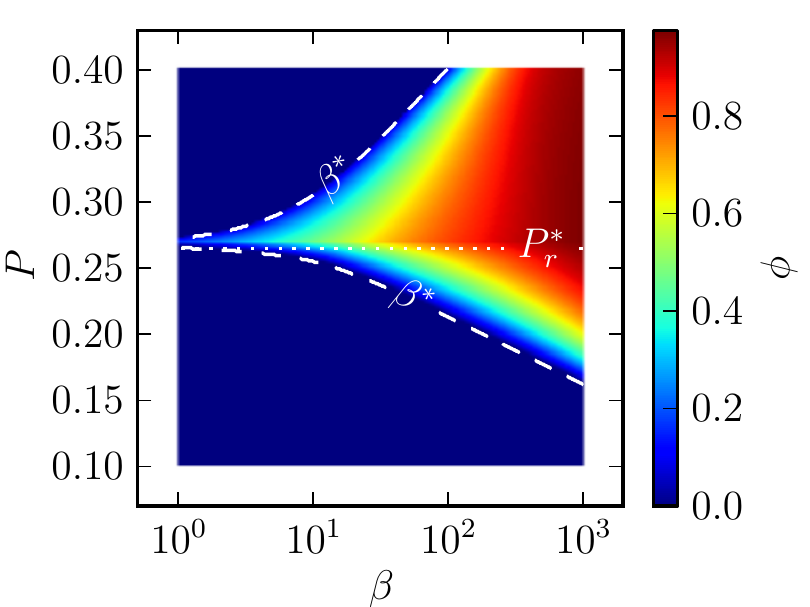}\\
      {\smaller[2] $\avg{k} = 7$}
    \end{center}
  \end{minipage}
  \begin{minipage}{0.49\columnwidth}
    \begin{center}
      \includegraphics*[width=1\columnwidth]{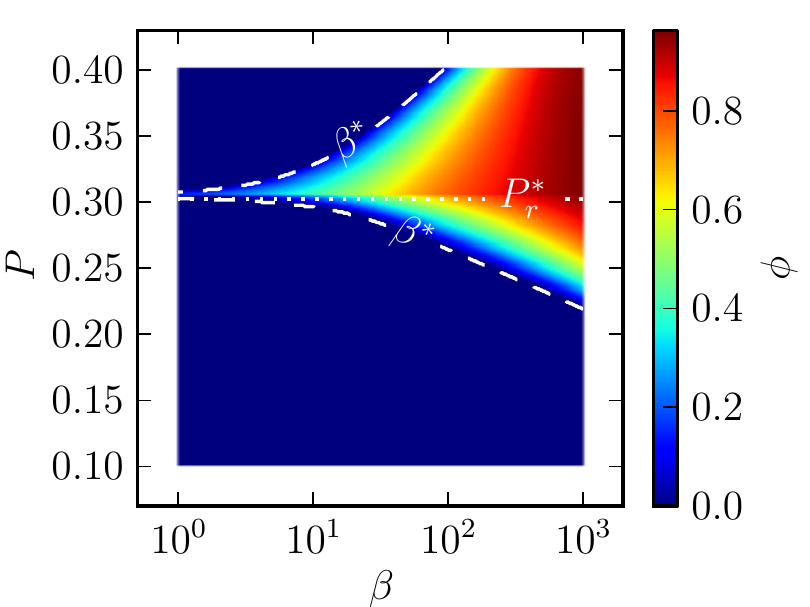}\\
      {\smaller[2] $\avg{k} = 10$}
    \end{center}
  \end{minipage}
  \caption{\label{fig:diag-phi}(Color online) The order parameter $\phi$ as a function
    of the selective pressure $\beta$ and noise $P$, for different
    values of $\avg{k}$.}
\end{figure}

\begin{figure} 
  \flushleft
  \includegraphics*[width=0.49\columnwidth]{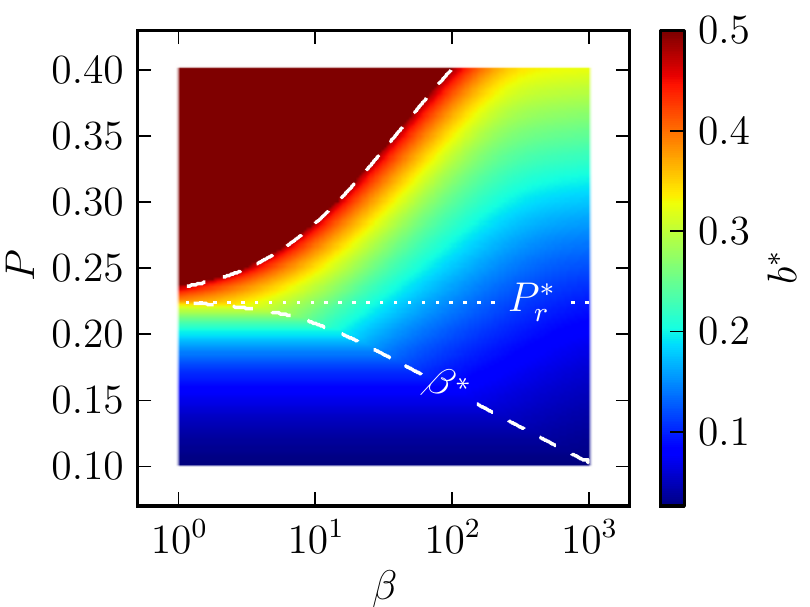}
  \includegraphics*[width=0.49\columnwidth]{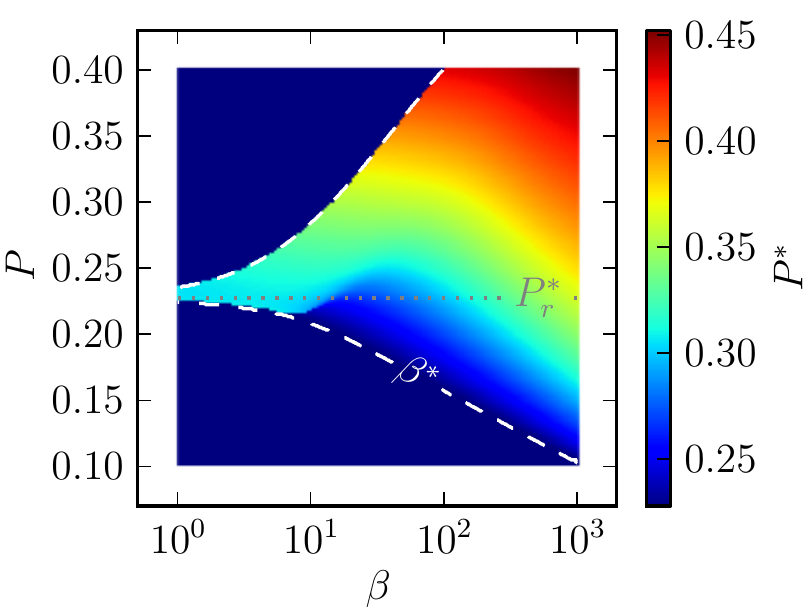}
  \caption{\label{fig:diag-pc}(Color online) \emph{Left:} Value of the steady-state
    average error $b^*$ as a function of the selective pressure $\beta$
    and noise $P$, for $\avg{k}=5$. \emph{Right:} Maximum tolerable
    noise $P^*$, as a function of the selective pressure $\beta$ and
    noise $P$, for $\avg{k}=5$.}
\end{figure}

As can be seen in Fig.~\ref{fig:diag-op}, there is a second-order phase
transition at a critical value $\beta^*$, which depends on the noise
level $P$. The dependence of $\beta^*$ on $P$ divides the $\beta\times
P$ phase diagram into an upper and lower branch, as can be seen in
Fig.~\ref{fig:diag-phi}. The branches are divided at a value of
$P=P^*_r$, which is the critical value of noise of a fully random
network (see~\cite{peixoto_behavior_2012} for an exact calculation of
$P^*_r$). At this value of noise, a random network undergoes a dynamic
phase transition, where the steady state error fraction reaches the
maximum level $b^*=1/2$, and the dynamics become ergodic, as was
described previously. For $ P < P^*_r$, random networks are
intrinsically robust, since $b^* < 1/2$, and the critical value
$\beta^*$ becomes larger with smaller $P$. In other words, the smaller
is the value of noise $P$, the better is the behavior of fully random
networks, such that the entropic cost of providing further improvement
by creating a segregated core becomes larger, which therefore occurs
only at larger values of selective pressure. The situation changes for
$P>P^*_r$. Since random networks are no longer resilient, and have
collapsed onto $b^*=1/2$ (see Fig.~\ref{fig:diag-pc}), a segregated core
provides a significant improvement, for a relatively low entropic
cost. This cost increases with $P$, since the core needs to be either
denser, smaller or more isolated to provide the same benefit under
larger noise. Thus the value of $\beta^*$ also increases with
$P$. Interestingly, in the vicinity of $P=P^*_r$, the value of $\beta^*$
tends to zero. For this value of noise, the dynamics of the fully random
topology lies exactly at the critical point where $b^*=1/2$, and even
the smallest (structural) perturbation can move the fixed point
appreciably. Since such small structural perturbations have negligible
entropic cost, the value of $\beta^*$ vanishes to zero. Thus, networks
with $\avg{k}$ such that $P^*_r=P$ are the most easily evolvable.

The value of $b^*$ itself can be seen in Fig.~\ref{fig:diag-pc}. The
upper branch $P>P^*_r$ corresponds to transitions from $b^*=1/2$
(ergodic dynamics) to $b^*<1/2$ (nonergodic dynamics), whereas the lower
branch $P<P^*_r$ shows $b^*<1/2$ for both phases.

\begin{figure} 
  \flushleft
  \begin{minipage}{\columnwidth}
    \includegraphics*[width=0.49\textwidth]{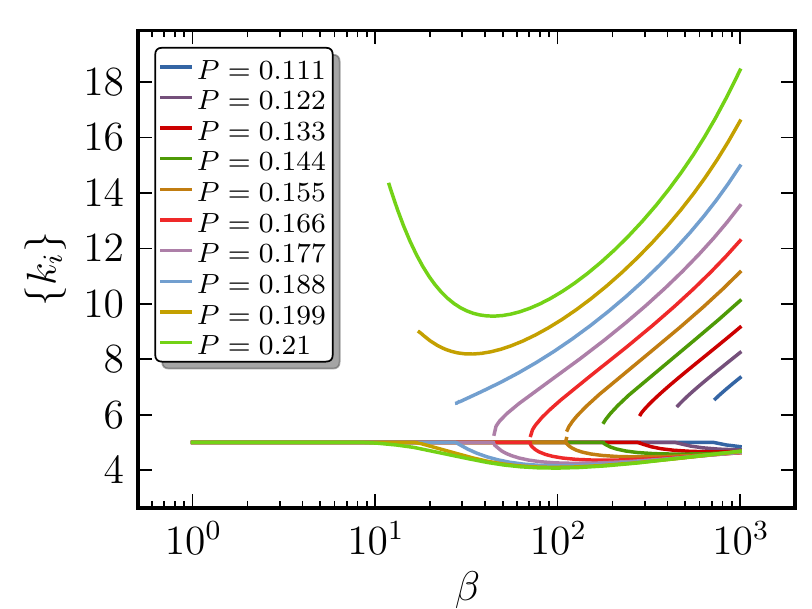}
    \includegraphics*[width=0.49\textwidth]{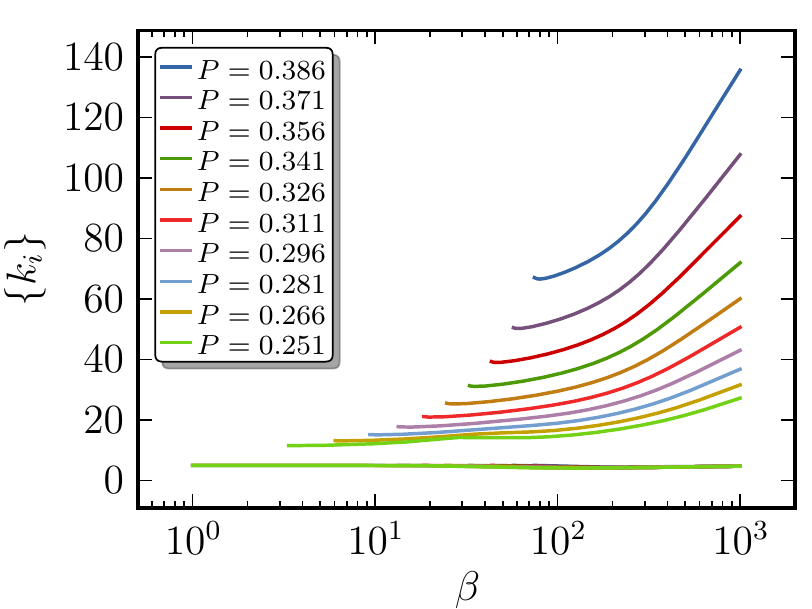}\\
    \includegraphics*[width=0.49\textwidth]{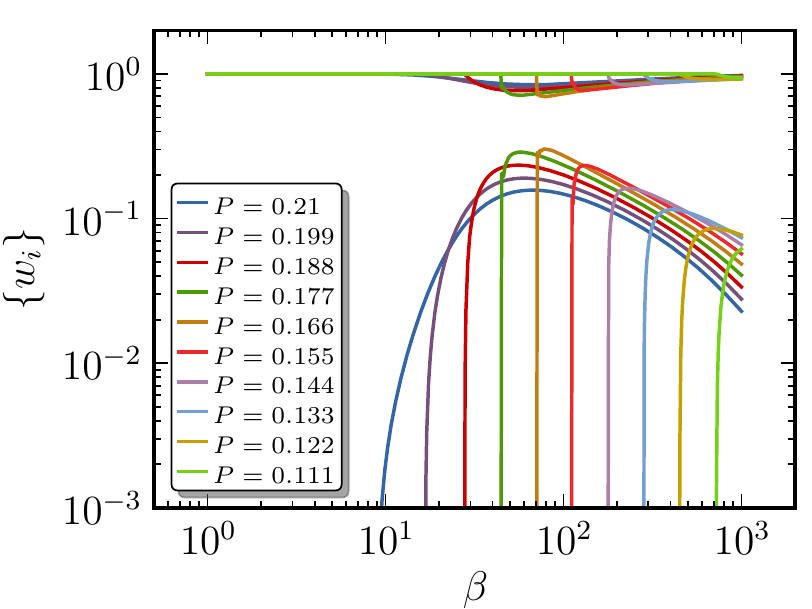}
    \includegraphics*[width=0.49\textwidth]{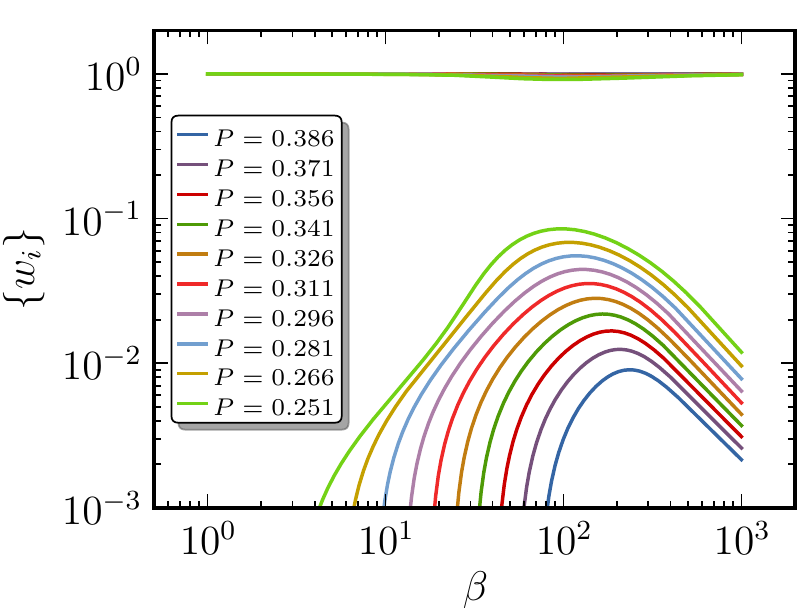}\\
    \includegraphics*[width=0.49\textwidth]{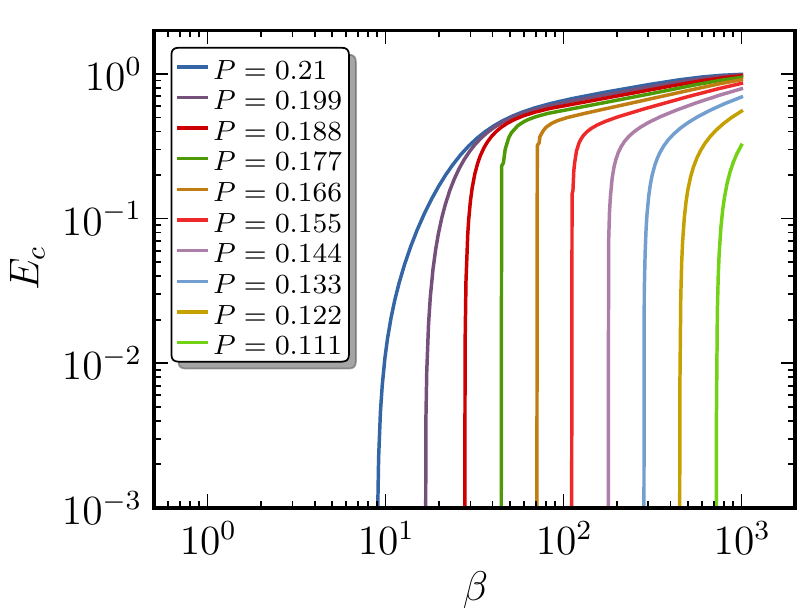}
    \includegraphics*[width=0.49\textwidth]{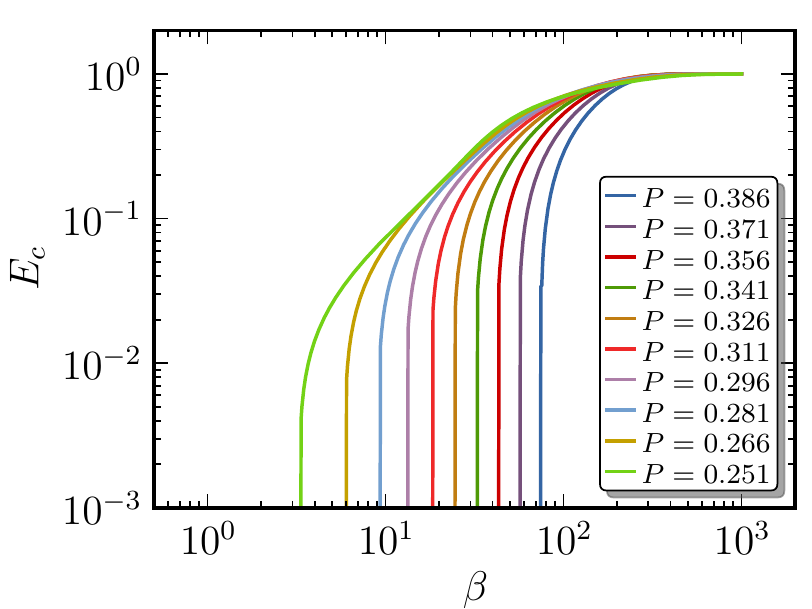}
  \end{minipage}
  \caption{\label{fig:diag-top}(Color online) Block average in-degrees $\{k_i\}$,
    block sizes $\{w_i\}$, and total fraction $E_c$ of inputs
    originating from the core block, as functions of selective pressure
    $\beta$, for $\avg{k} = 5$. The left panels show curves for $P <
    P^*_r$, where $P^*_r$ is the critical value of noise for a fully
    random network, and the right panels show curves for $P >
    P^*_r$. All curves on the left panels are shown in order of
    increasing $P$ from right to left, and on the right panel, from left
    to right.}
\end{figure}

The topological properties of each phase can be seen in detail in
Fig.~\ref{fig:diag-top}, where are shown the average in-degrees
$\{k_i\}$, block sizes $\{w_i\}$ and the total fraction of inputs which
lead to the segregated core, $E_c=\sum_j w_{c\to j} w_j k_j / \avg{k}$,
where $c$ is the core block. The core block emerges at $\beta =
\beta^*$, with an infinitesimal size, but with a value of $k_i$ which is
usually significantly above average. For $P>P^*_r$ the segregated core
usually has a significantly larger average in-degree than for
$P<P^*_r$. The dominance and segregation of the core block increases
continuously with $\beta$, reaching values close to $E_c=1$, for larger
values of $\beta$, especially for values of $P>P^*_r$.

Each network on the evolved ensemble has a critical value of noise $P^*$
(different from the value of $P$ for which it was evolved), for which
its dynamics undergoes the aforementioned ergodicity transition and
which represents the maximum tolerable noise
(see~\cite{peixoto_behavior_2012} for an exact calculation of $P^*$ for
arbitrary blockmodels). Interestingly, the evolution of $b^*$ does not
automatically result in larger values of $P^*$, as is shown in
Fig.~\ref{fig:diag-pc}: Some ensembles evolved under larger selective
pressure possess a lower value of $P^*$ than others evolved under lower
selective pressure (for the same value of $P$ under evolution). This
means the evolution is reasonably specialized for the level of noise it
is under, and the behavior of the networks under larger values of noise
for which they were evolved is not automatically better than that of
other networks with smaller fitness. However, despite these deviations,
the general tendency is that, for larger values of $\beta$, $b^*$ and
$P^*$ are decreased and increased, respectively.

\section{Monte-Carlo simulations}\label{sec:mc}

We have also performed Monte Carlo simulations to observe the
phase-transition obtained in the previous section. We employed the
Metropolis-Hastings~\cite{metropolis_equation_1953, hastings_monte_1970}
algorithm, starting from a random network with $N$ vertices, with a
given average in-degree $\avg{k}$ and a partition into $n$ blocks,
represented by assigning block labels to each vertex (which is initially
randomly chosen). At each iteration, one of the following moves is
attempted with equal probability:
\begin{enumerate}
\item \emph{Block label move}: A random vertex $v$ is chosen, and its
  block label is randomly chosen among all $n$ possible values. \label{it:label}
\item \emph{Input move}: A vertex $v$ is chosen with probability
  $p\propto k(k-1)$, where $k$ is the in-degree of $v$. Another vertex
  $u$ is randomly chosen with uniform probability. Two random inputs
  from $v$ are deleted and moved to $u$. \label{it:input}
\item \emph{Source move}: A random vertex $v$ is chosen. A random input
  from $v$ is deleted and replaced by a randomly chosen one.
\end{enumerate}
A move is rejected if it generates parallel edges or self-loops. The
difference $\Delta b^*$ of the value of $b^*$ after and before the move
is computed. The move is then accepted with a probability $p_a$ given by
\begin{equation}
  p_a =
  \begin{cases}
    1 &\text{ if } \Delta b^* \le 0,\\
    e^{-\beta N\Delta b^*} &\text{ otherwise.}
  \end{cases}
\end{equation}
The probability $p\propto k(k-1)$ in move~(\ref{it:input}) is chosen to
correspond to \emph{two} independent single-edge moves affecting the
same vertices $v$ and $u$, where in each move a random edge is chosen,
and its target is moved to a randomly chosen vertex. This guarantees
that there is no topological bias, and that the in-degrees are always
odd.

\begin{figure} 
  \flushleft
  \includegraphics*[width=0.49\columnwidth]{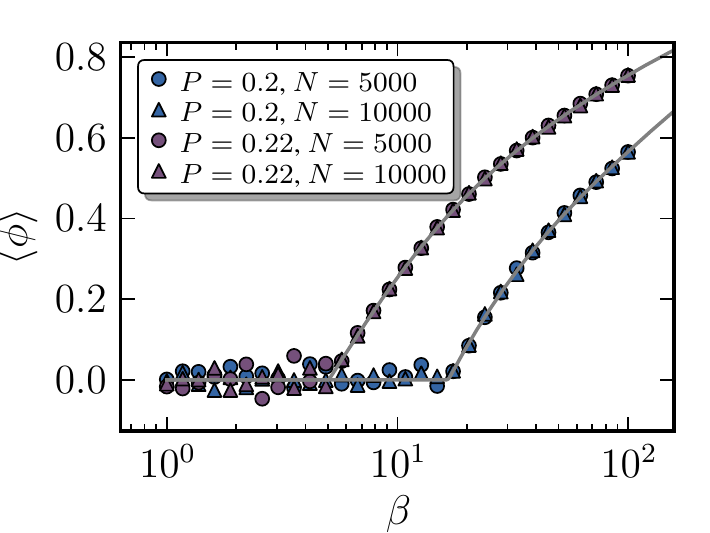}
  \includegraphics*[width=0.49\columnwidth]{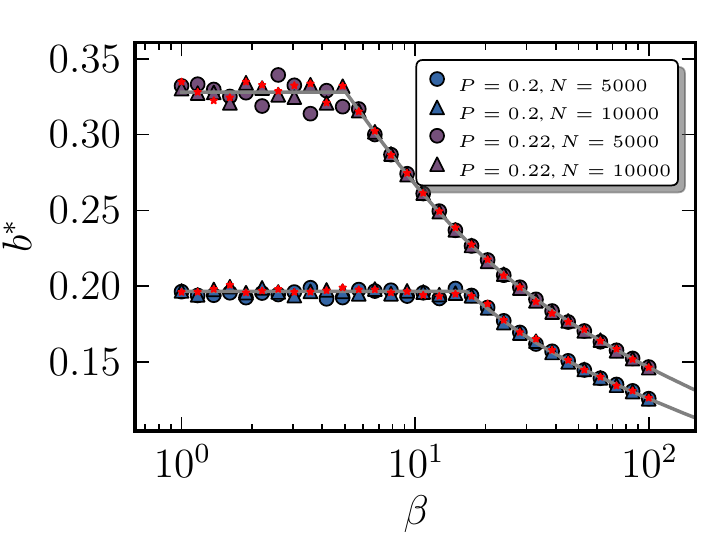}
  \caption{\label{fig:mc-trans}(Color online) Average order parameter $\avg{\phi}$ and
    steady-state error level $b^*$ as functions of the selective
    pressure $\beta$, for different values of noise $P$ and $\avg{k}=5$,
    obtained with Monte Carlo simulations, for network sizes shown in
    the legend. On the right plot, the red star symbols ($\star$)
    correspond to empirical values of $b^*$ as obtained with
    Eq.~\ref{eq:bn_dyn}. The solid gray lines are theoretical values
    obtained by minimizing the free energy.}
\end{figure}

\begin{figure} 
  \flushleft
  \includegraphics*[width=0.49\columnwidth]{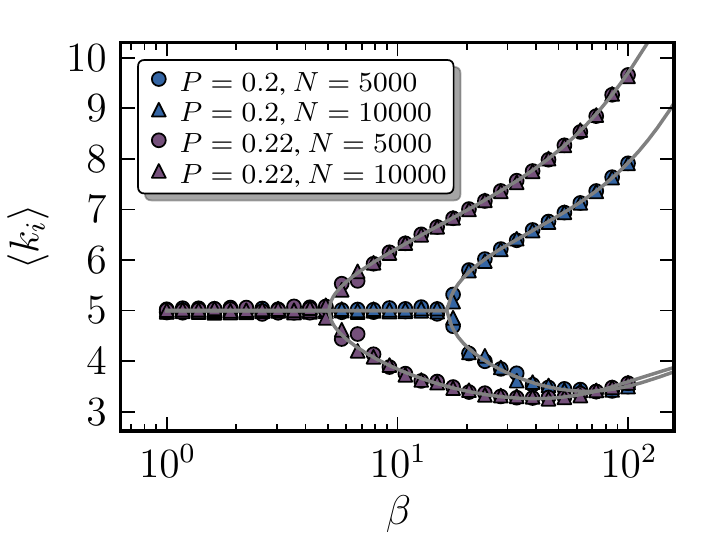}
  \includegraphics*[width=0.49\columnwidth]{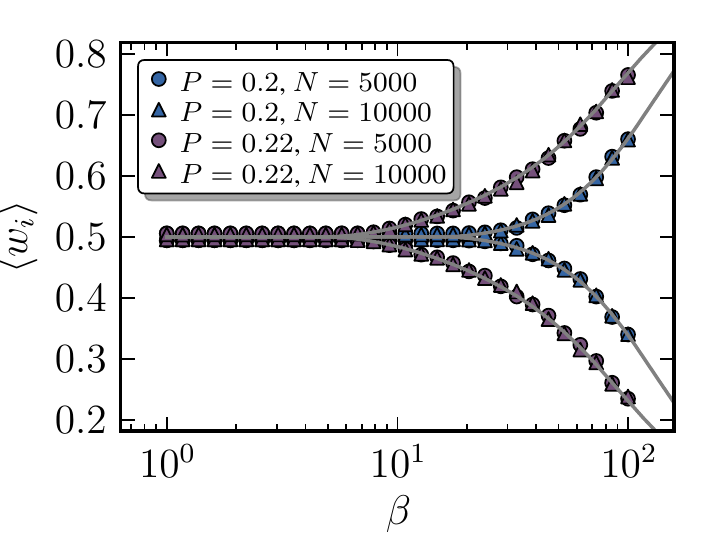}
  \caption{\label{fig:mc-top}(Color online) Block average in-degrees $\{k_i\}$ and
    sizes $\{w_i\}$, as functions of selective pressure $\beta$, for
    $\avg{k} = 5$, obtained with Monte-Carlo simulations, for network
    sizes shown in the legend. The solid gray lines are theoretical
    values obtained by minimizing the free energy.}
\end{figure}

\begin{figure} 
  \flushleft
  \includegraphics*[width=0.49\columnwidth]{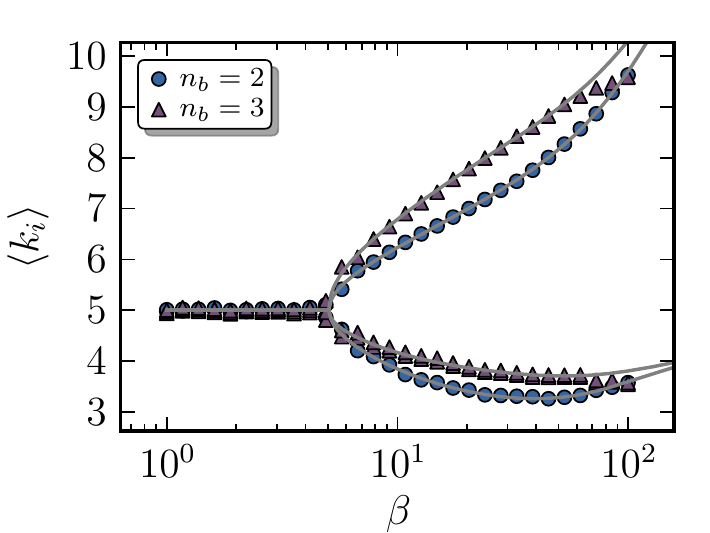}
  \includegraphics*[width=0.49\columnwidth]{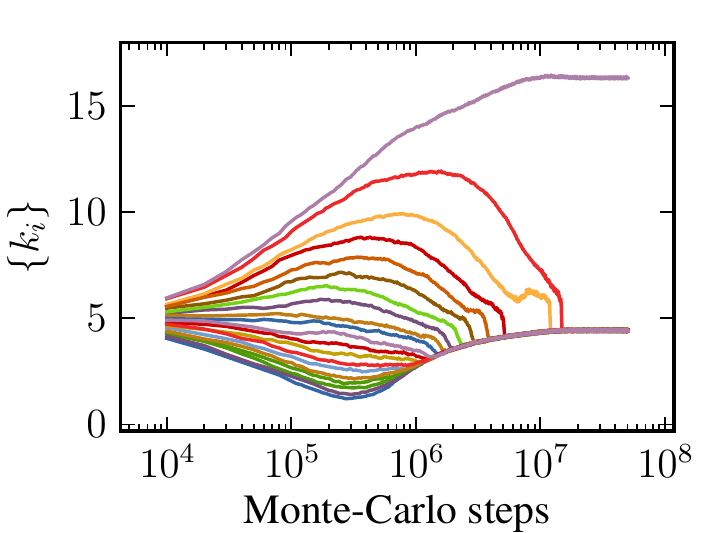}
  \caption{\label{fig:mc-multiblock}(Color online) \emph{Left:} Block average
    in-degrees $\{k_i\}$, as a function of selective pressure $\beta$,
    for $\avg{k} = 5$, and obtained with Monte-Carlo simulations, for
    $N=10000$ and different number of blocks. The solid gray lines are
    theoretical values obtained by minimizing the free energy.
    \emph{Right:} Evolution in time of the average in-degrees $\{k_i\}$
    in a Monte-Carlo simulation with $n=20$ blocks, $N=10000$ and
    $\beta=10^3$, showing the eventual merging into only two blocks.}
\end{figure}

The value of $b^*$ is computed by obtaining the values of $\{w_i\}$,
$\{w_{j\to i}\}$ and $\{k_i\}$, and iterating
Eq.~\ref{eq:dynblock}. This is much faster than actually measuring the
error level on the network, and produces deterministic
values~\footnote{One could argue that this may overlook the buildup of
  correlations in the network, since it assumes that the blocks are
  homogeneous and have an random in-degree distribution as given by
  Eq.~\ref{eq:pk}. However, we are \emph{only} interested in networks
  which have this property, which, as discussed in the text, correspond
  to a partial maximization of entropy when the remaining constraints
  are in place, so this should not be an issue. To be sure, we have
  compared the value of $b^*$ computed this way with the actual
  empirical value and found a very good agreement (see
  Fig.~\ref{fig:mc-trans}).}.

Since we have employed the block label move~(\ref{it:label}), which
tends to partition the network evenly into $n$ blocks of equal sizes, we
have included an entropic cost associated with the size of a block,
which did not exist in the original blockmodel above. In the original
model, the partitions themselves are not relevant, and only the
resulting graph topology contributes to the entropy. However,
move~(\ref{it:label}) makes the algorithm very efficient and easy to
implement, and it should not fundamentally change the results. But in
order to compare with the theory, we need to include the following
correction in the number of possible networks:
\begin{equation}
  \Omega' = \Omega \times \frac{N!}{\prod_i(N w_i)!}
\end{equation}
which leads to the slightly modified entropy,
\begin{equation}\label{eq:part}
  \mathcal{S}'/N =   \mathcal{S}/N - \sum_iw_i\ln w_i.
\end{equation}

In Fig.~\ref{fig:mc-trans} we can see the same phase transition observed
previously, which matches very well the theoretical predictions. In
Fig.~\ref{fig:mc-top} the topology can be assessed more closely, and the
emergence of the segregated core is clear. Due to the partition entropy
introduced in Eq.~\ref{eq:part}, the core does not vanish at the
transition; it merges continuously with the other block
instead. However, the critical value $\beta^*$ is identical with the
non-modified model.

The inclusion of the partition entropy also introduces the fact that
different solutions are obtained for a different number of blocks, since
this has a direct effect on the preferred sizes of the blocks (see
Fig.~\ref{fig:mc-multiblock}, left). However, this \emph{does not}
change the fact that for any number of blocks the preferred topology
will always be an effective two-block structure. This follows from the
argumentation presented previously based on the reduction of entropy
resulting from block splits, and can be observed in simulations with
many blocks, as shown in Fig.~\ref{fig:mc-top}, which shows a comparison
between the topologies obtained with two and three blocks, as well as
the outcome of a typical simulation with $20$ blocks, which shows the
eventual collapse of into an effective two-block structure.

\section{Gene regulatory networks}\label{sec:real}

Here we make a comparison with some features observed in actual gene
regulatory networks. We consider the networks for \emph{Saccharomyces
  cerevisiae} (yeast) and \emph{Escherichia coli}, extracted from the
YEASTRACT~\cite{abdulrehman_yeastract:_2010} and
RegulonDB~\cite{gama-castro_regulondb_2010} databases, respectively.  We
are interested in extracting the ``functional core'' of the network,
i.e. those nodes which are solely responsible for global regulation,
like those belonging to the \emph{segregated core} which emerges in the
phase transition observed in the evolutionary process above. We will
characterize the core nodes in two ways: 1. Nodes which have an
out-degree larger than zero; 2. Nodes which belong to a \emph{strongly
  connected component} (SCC) of the graph (i.e. the maximal set of nodes
which can directly or indirectly regulate each other). The first
criterion is a necessary condition, since if the out-degree is zero,
then a node is not a regulator. The second criterion is stronger, since
even if a node is a regulator, it can have its dynamics completely
enslaved by other nodes. The nodes in the SCC are exactly those which
are not necessarily enslaved, and can mutually regulate each
other. Without a least one SCC in the network, an autonomous behavior
with dynamical attractors other than simple fixed-points is not
possible.

\begin{figure} [tb]
  \includegraphics*[width=0.49\columnwidth]{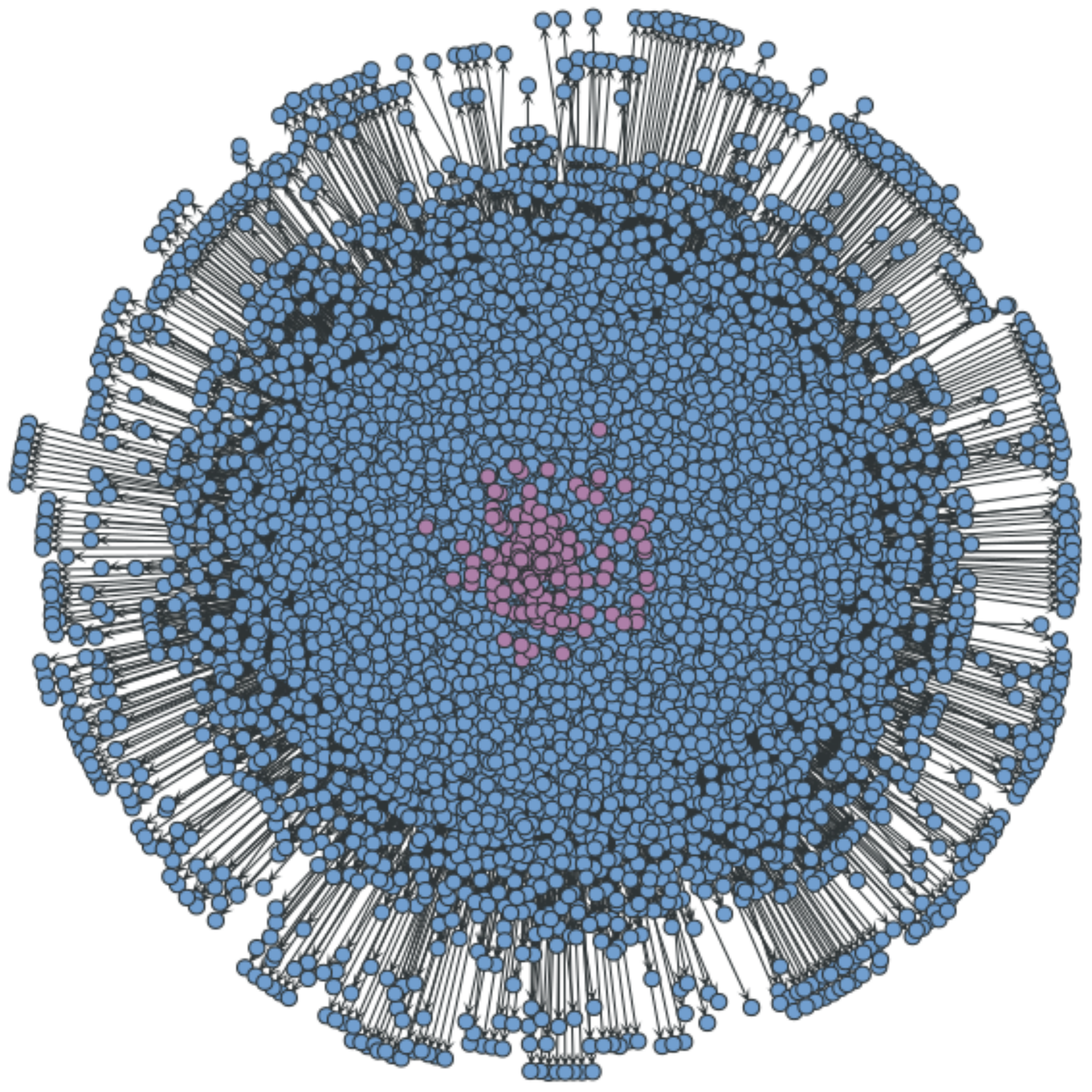}
  \includegraphics*[width=0.49\columnwidth]{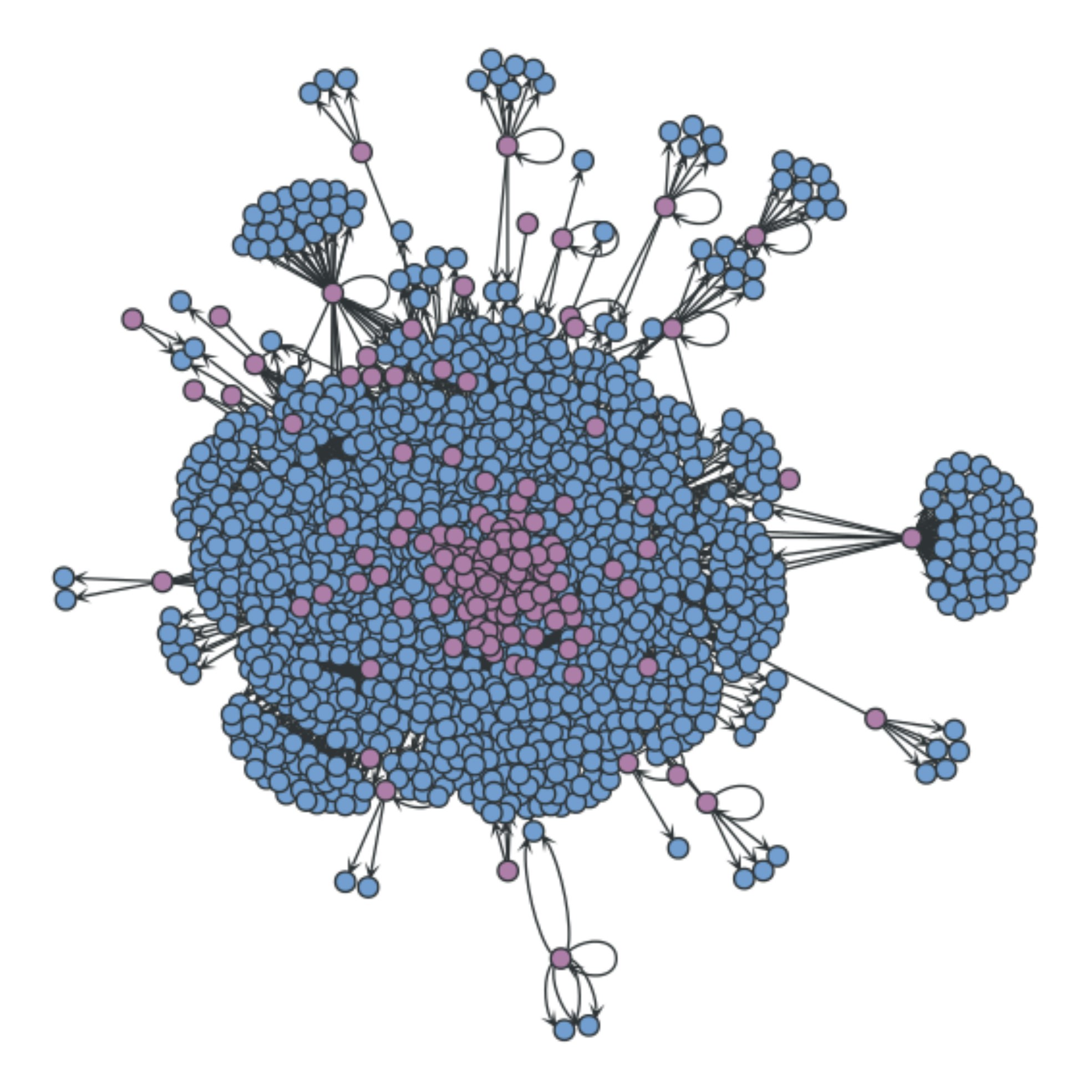}
  \caption{\label{fig:gene}(Color online) Gene regulatory networks for
    \emph{Saccharomyces cerevisiae} (left) and \emph{Escherichia coli}
    (right), extracted from the
    YEASTRACT~\cite{abdulrehman_yeastract:_2010} and
    RegulonDB~\cite{gama-castro_regulondb_2010} databases,
    respectively. The nodes in purple (towards the middle) are
    transcription factor genes, and are the only ones with out-degree
    larger than zero.}
\end{figure}

The yeast network is composed of $N=6402$ nodes, with an average
in-degree of $\avg{k} \approx 7.51$. The \emph{E. coli} network is
smaller and sparser, with $N=1658$ and $\avg{k} \approx 2.32$. In both
networks the number of transcription factor (TF) genes is much smaller
than the total number: $N_{\text{TF}} = 182$ in yeast, and
$N_{\text{TF}} = 154$ in \emph{E. coli}. These are core genes according
to the first criterion, since they have an out-degree larger than zero,
as can be seen in Fig.~\ref{fig:gene}.

In yeast, the average in-degree of the core nodes is higher than
average, $\avg{k}_c \approx 10.03$, as observed in the segregated core
phase of the evolved networks obtained. For the SCC, the number of nodes
decreases slightly to $N_{\text{SCC}}=146$, and the average in-degree
changes negligibly $\avg{k}_{CC} \approx 10.48$ (if one counts only
edges between vertices of the SCC, this value is virtually identical,
$\avg{k}_{CC} \approx 10.42$).  This is similar to what was found
previously in~\cite{maslov_computational_2005} for the yeast network
(using an older, and less complete dataset with only $837$ genes). They
have also found that the TF genes have different connection patterns,
and those with the largest out-degree tend to regulate genes with lower
than average in-degree. However they did not find that the TF genes form
a \emph{denser} subgraph, with an larger than average in-degree, which
is most likely due to the incompleteness of the dataset used. Very
similar numbers to those presented here were obtained more recently
in~\cite{balaji_comprehensive_2006}, using a more complete dataset
(which is not identical to the one used in this work).

For \emph{E. coli} the situation changes somewhat: The average in-degree
of the transcription factor nodes is $\avg{k}_c \approx 1.97$, which is
in fact \emph{lower} than the global average. However, if one extracts
the largest SCC, the number of nodes drops dramatically to
$N_{\text{SCC}}=8$. These nodes are responsible for the regulation of
$411$ genes. A majority of $1093$ genes are instead enslaved to the
dynamics of SCC with only two mutually regulating nodes. Although the
largest SCC \emph{does} have an average in-degree $\avg{k}_{CC} = 6$,
the core topology seems significantly more sparse than for yeast and the
evolved networks, at least with the data currently
available~\cite{cosentino_lagomarsino_hierarchy_2007}. Arguably, such a
sparse regulating core is suspect from the point of view of data set
completeness, since it would mean that the range of dynamical behavior
for the regulatory network is very restricted. As previously mentioned,
and older and less complete data set for yeast also did not reveal a
denser regulating core~\cite{maslov_computational_2005}. Nevertheless,
one should also consider that such real networks are simultaneously
under different, possibly competing selective pressures which also
influence the resulting topology, robustness against noise being only
one of them. These other factors, which are neglected in the model,
could be one reason for such a discrepancy. We emphasize, however, that
although apparently it is not denser, a regulating core certainly exists
in the measured \emph{E. coli} network, which is at least in partial
qualitative agreement with what is observed in the model.

There are other factors that may contribute to this observed segregation
which are not in principle related to noise resilience. For instance,
non-regulating genes exist mostly to transcribe proteins which have some
specific metabolic or structural function in the cell, and it may be
difficult for these proteins to have a dual role as transcription
factors, and therefore become specialized (although non-specialized
proteins are not impossible, since a protein can in principle bind both
to DNA and to other proteins). Nevertheless, there are good reasons to
consider robustness to noise as a very plausible driving force toward
this type of topology.  This is corroborated, for instance, by evidence
that core TF genes tend to be less noisy~\cite{fraser_noise_2004,
jothi_genomic_2009}, and that the vast majority of TF genes in yeast are
not vital for the survival of the cell if repressed in
isolation~\cite{balaji_uncovering_2006}. This is fully compatible with
the idea of a highly redundant functional core, which provides
robustness for the rest of the network.

Another feature which is commonly investigated in empirical networks is
the in- and out-degree distributions. The in-degree distribution is
often narrow, while the out-degree distribution is broader, and as some
suggest~\cite{maslov_computational_2005}, compatible with a power
law. The model considered in this work is parametrized as a stochastic
blockmodel, where each block has in- and out-degrees that are Poisson
distributed. When the segregated core emerges, the system is composed of
only two blocks, thus both the in- and out-degree distributions are
bimodal. The in-degree distribution is indeed narrower, since the
difference between the average in-degree of the two blocks is not very
large for most networks obtained. The out-degree distribution is also
much broader, since the average out-degree of the non-core block tends
to zero, while for the core block it tends very rapidly to infinity,
when the selective pressure is increased. However, the homogeneous and
seemingly scale-free properties of the empirical distributions are not
reproduced by the model. This implies that these features are not a
direct result of evolved robustness against noise, and may, for
instance, be due simply to mutational bias caused by gene duplication,
which is known to qualitatively reproduce these types of in- and
out-degree distributions~\cite{ispolatov_duplication-divergence_2005,
  enemark_gene_2007}.

\section{Conclusion} \label{sec:conclusion}

We have investigated the effect of selective pressure favoring
robustness against noise on the structural evolution of Boolean networks
with optimal majority functions, functioning as a conceptual model for
gene regulation. We have mapped the evolutionary process onto a Gibbs
ensemble, and obtained its outcome by minimizing the associated free
energy. We showed that the structural properties of the system undergo a
phase transition at a critical value of selective pressure, from a
random topology to a segregated-core structure, where a smaller fraction
of the nodes form an isolated core, which is denser than the rest of the
network and is responsible for most of the regulation. Since the core is
denser, its nodes can profit from more regulatory redundancy, which
greatly diminishes the effect of noise. This robustness is propagated to
the rest of the network, which relies on the core for most of the
regulation.  The segregated core becomes denser, smaller, and more
isolated as the selective pressure increases. We have compared the
theoretical predictions with Monte-Carlo simulations of actual networks,
and found perfect agreement.

We have also shown that this segregated-core structure is present in the
gene regulatory network of yeast and \emph{E. coli}. Both networks are
composed of a much smaller fraction of transcription factor genes which
are responsible for all regulation. In yeast, the existing core
structure is very similar qualitatively to the outcome of the
evolutionary process considered, with transcription factor genes forming
a denser subgraph, with an average in-degree above the average for the
whole network. In \emph{E. coli} the isolated transcription factor core
is composed of few very small regulating cores (strongly connected
components), the largest of which has only eight nodes. We conjecture
that such a sparse regulating core is possibly due to data set
incompleteness, since it would severely restrict the range of possible
dynamical behavior for the network. A less complete data set for yeast
also did not show a denser regulating core
structure~\cite{maslov_computational_2005}, although it is clearly seen
with more up-to-date datasets including more genes and
interactions~\cite{balaji_comprehensive_2006}. However, one should not
rule out other selection criteria which are not incorporated in the
model.

It should also be noted that regulating cores of transcription factors
are a common feature of other organisms, such as \emph{Mycobacterium
  tuberculosis}~\cite{balazsi_temporal_2008}. Additionally, a
similar (but not identical) ``bow-tie'' structure was also observed in
the mammalian signal transduction
network~\cite{maayan_formation_2005,maayan_insights_2009}, where most
pathways are funneled through a central core.

It is possible to formulate other reasons for the existence of such a
core structure, such as a forced specialization of genes into either
transcription factors or target genes. Furthermore one should mention
that the effects of noise are not always detrimental, and can in some
circumstances even be beneficial~\cite{blake_phenotypic_2006,
  eldar_functional_2010}. Nevertheless there is compelling evidence
that the core genes provide a degree of robustness to the cell. Not only
are the best-connected TF nodes less noisy~\cite{fraser_noise_2004,
  jothi_genomic_2009}, they are usually found --- if removed
individually --- not to be vital for cell
survival~\cite{balaji_uncovering_2006}. This corroborates the idea that
one of the major functions of the regulating core is to provide
robustness via redundancy.

Furthermore, aside from the direct applicability to gene regulation, we
have identified a fundamental mechanism of robustness against noise,
which emerges naturally when networks are randomly selected for that
purpose~\footnote{Other evolutionary, non-equilibrium pathways are also
possible, see e.g.~\cite{perotti_emergent_2009}.}. Although most
interesting systems require more than just robustness for their
functioning, it is reasonable to conclude that the emergence of
regulating cores is to be expected when there is enough selective
pressure favoring noise resilience.

\bibliographystyle{apsrev4-1}
\bibliography{bib}

\begin{thebibliography}{69}%
\makeatletter
\providecommand \@ifxundefined [1]{%
 \@ifx{#1\undefined}
}%
\providecommand \@ifnum [1]{%
 \ifnum #1\expandafter \@firstoftwo
 \else \expandafter \@secondoftwo
 \fi
}%
\providecommand \@ifx [1]{%
 \ifx #1\expandafter \@firstoftwo
 \else \expandafter \@secondoftwo
 \fi
}%
\providecommand \natexlab [1]{#1}%
\providecommand \enquote  [1]{``#1''}%
\providecommand \bibnamefont  [1]{#1}%
\providecommand \bibfnamefont [1]{#1}%
\providecommand \citenamefont [1]{#1}%
\providecommand \href@noop [0]{\@secondoftwo}%
\providecommand \href [0]{\begingroup \@sanitize@url \@href}%
\providecommand \@href[1]{\@@startlink{#1}\@@href}%
\providecommand \@@href[1]{\endgroup#1\@@endlink}%
\providecommand \@sanitize@url [0]{\catcode `\\12\catcode `\$12\catcode
  `\&12\catcode `\#12\catcode `\^12\catcode `\_12\catcode `\%12\relax}%
\providecommand \@@startlink[1]{}%
\providecommand \@@endlink[0]{}%
\providecommand \url  [0]{\begingroup\@sanitize@url \@url }%
\providecommand \@url [1]{\endgroup\@href {#1}{\urlprefix }}%
\providecommand \urlprefix  [0]{URL }%
\providecommand \Eprint [0]{\href }%
\providecommand \doibase [0]{http://dx.doi.org/}%
\providecommand \selectlanguage [0]{\@gobble}%
\providecommand \bibinfo  [0]{\@secondoftwo}%
\providecommand \bibfield  [0]{\@secondoftwo}%
\providecommand \translation [1]{[#1]}%
\providecommand \BibitemOpen [0]{}%
\providecommand \bibitemStop [0]{}%
\providecommand \bibitemNoStop [0]{.\EOS\space}%
\providecommand \EOS [0]{\spacefactor3000\relax}%
\providecommand \BibitemShut  [1]{\csname bibitem#1\endcsname}%
\let\auto@bib@innerbib\@empty
\bibitem [{\citenamefont {Kitano}(2004)}]{kitano_biological_2004}%
  \BibitemOpen
  \bibfield  {author} {\bibinfo {author} {\bibfnamefont {H.}~\bibnamefont
  {Kitano}},\ }\href {\doibase 10.1038/nrg1471} {\bibfield  {journal} {\bibinfo
   {journal} {Nat Rev Genet}\ }\textbf {\bibinfo {volume} {5}},\ \bibinfo
  {pages} {826} (\bibinfo {year} {2004})}\BibitemShut {NoStop}%
\bibitem [{\citenamefont {Raser}\ and\ \citenamefont
  {{O'Shea}}(2005)}]{raser_noise_2005}%
  \BibitemOpen
  \bibfield  {author} {\bibinfo {author} {\bibfnamefont {J.~M.}\ \bibnamefont
  {Raser}}\ and\ \bibinfo {author} {\bibfnamefont {E.~K.}\ \bibnamefont
  {{O'Shea}}},\ }\href {\doibase 10.1126/science.1105891} {\bibfield  {journal}
  {\bibinfo  {journal} {Science}\ }\textbf {\bibinfo {volume} {309}},\ \bibinfo
  {pages} {2010 } (\bibinfo {year} {2005})}\BibitemShut {NoStop}%
\bibitem [{\citenamefont {Maheshri}\ and\ \citenamefont
  {{O’Shea}}(2007)}]{maheshri_living_2007}%
  \BibitemOpen
  \bibfield  {author} {\bibinfo {author} {\bibfnamefont {N.}~\bibnamefont
  {Maheshri}}\ and\ \bibinfo {author} {\bibfnamefont {E.~K.}\ \bibnamefont
  {{O’Shea}}},\ }\href {\doibase 10.1146/annurev.biophys.36.040306.132705}
  {\bibfield  {journal} {\bibinfo  {journal} {Annual Review of Biophysics and
  Biomolecular Structure}\ }\textbf {\bibinfo {volume} {36}},\ \bibinfo {pages}
  {413} (\bibinfo {year} {2007})}\BibitemShut {NoStop}%
\bibitem [{\citenamefont {Gutkind}(2000)}]{gutkind_signaling_2000}%
  \BibitemOpen
  \bibfield  {author} {\bibinfo {author} {\bibfnamefont {J.~S.}\ \bibnamefont
  {Gutkind}},\ }\href@noop {} {\emph {\bibinfo {title} {Signaling networks and
  cell cycle control: the molecular basis of cancer and other diseases}}}\
  (\bibinfo  {publisher} {Humana Press},\ \bibinfo {year} {2000})\BibitemShut
  {NoStop}%
\bibitem [{\citenamefont {Lestas}\ \emph {et~al.}(2010)\citenamefont {Lestas},
  \citenamefont {Vinnicombe},\ and\ \citenamefont
  {Paulsson}}]{lestas_fundamental_2010}%
  \BibitemOpen
  \bibfield  {author} {\bibinfo {author} {\bibfnamefont {I.}~\bibnamefont
  {Lestas}}, \bibinfo {author} {\bibfnamefont {G.}~\bibnamefont {Vinnicombe}},
  \ and\ \bibinfo {author} {\bibfnamefont {J.}~\bibnamefont {Paulsson}},\
  }\href {\doibase 10.1038/nature09333} {\bibfield  {journal} {\bibinfo
  {journal} {Nature}\ }\textbf {\bibinfo {volume} {467}},\ \bibinfo {pages}
  {174} (\bibinfo {year} {2010})}\BibitemShut {NoStop}%
\bibitem [{\citenamefont
  {Kauffman}(1969{\natexlab{a}})}]{kauffman_metabolic_1969}%
  \BibitemOpen
  \bibfield  {author} {\bibinfo {author} {\bibfnamefont {S.~A.}\ \bibnamefont
  {Kauffman}},\ }\href {\doibase 10.1016/0022-5193(69)90015-0} {\bibfield
  {journal} {\bibinfo  {journal} {Journal of Theoretical Biology}\ }\textbf
  {\bibinfo {volume} {22}},\ \bibinfo {pages} {437} (\bibinfo {year}
  {1969}{\natexlab{a}})}\BibitemShut {NoStop}%
\bibitem [{\citenamefont
  {Kauffman}(1969{\natexlab{b}})}]{kauffman_homeostasis_1969}%
  \BibitemOpen
  \bibfield  {author} {\bibinfo {author} {\bibfnamefont {S.}~\bibnamefont
  {Kauffman}},\ }\href {\doibase 10.1038/224177a0} {\bibfield  {journal}
  {\bibinfo  {journal} {Nature}\ }\textbf {\bibinfo {volume} {224}},\ \bibinfo
  {pages} {177} (\bibinfo {year} {1969}{\natexlab{b}})}\BibitemShut {NoStop}%
\bibitem [{\citenamefont {Bornholdt}(2005)}]{bornholdt_systems_2005}%
  \BibitemOpen
  \bibfield  {author} {\bibinfo {author} {\bibfnamefont {S.}~\bibnamefont
  {Bornholdt}},\ }\href {\doibase 10.1126/science.1119959} {\bibfield
  {journal} {\bibinfo  {journal} {Science}\ }\textbf {\bibinfo {volume}
  {310}},\ \bibinfo {pages} {449} (\bibinfo {year} {2005})}\BibitemShut
  {NoStop}%
\bibitem [{\citenamefont {Drossel}(2008)}]{drossel_random_2008}%
  \BibitemOpen
  \bibfield  {author} {\bibinfo {author} {\bibfnamefont {B.}~\bibnamefont
  {Drossel}},\ }in\ \href {http://arxiv.org/abs/0706.3351} {\emph {\bibinfo
  {booktitle} {Reviews of Nonlinear Dynamics and Complexity}}},\ Vol.~\bibinfo
  {volume} {1},\ \bibinfo {editor} {edited by\ \bibinfo {editor} {\bibfnamefont
  {H.~G.}\ \bibnamefont {Schuster}}}\ (\bibinfo  {publisher} {Wiley},\ \bibinfo
  {year} {2008})\BibitemShut {NoStop}%
\bibitem [{\citenamefont {Raj}\ and\ \citenamefont {van
  Oudenaarden}(2008)}]{raj_nature_2008}%
  \BibitemOpen
  \bibfield  {author} {\bibinfo {author} {\bibfnamefont {A.}~\bibnamefont
  {Raj}}\ and\ \bibinfo {author} {\bibfnamefont {A.}~\bibnamefont {van
  Oudenaarden}},\ }\href {\doibase 16/j.cell.2008.09.050} {\bibfield  {journal}
  {\bibinfo  {journal} {Cell}\ }\textbf {\bibinfo {volume} {135}},\ \bibinfo
  {pages} {216} (\bibinfo {year} {2008})}\BibitemShut {NoStop}%
\bibitem [{\citenamefont {Eldar}\ and\ \citenamefont
  {Elowitz}(2010)}]{eldar_functional_2010}%
  \BibitemOpen
  \bibfield  {author} {\bibinfo {author} {\bibfnamefont {A.}~\bibnamefont
  {Eldar}}\ and\ \bibinfo {author} {\bibfnamefont {M.~B.}\ \bibnamefont
  {Elowitz}},\ }\href {\doibase 10.1038/nature09326} {\bibfield  {journal}
  {\bibinfo  {journal} {Nature}\ }\textbf {\bibinfo {volume} {467}},\ \bibinfo
  {pages} {167} (\bibinfo {year} {2010})}\BibitemShut {NoStop}%
\bibitem [{\citenamefont {Kollmann}\ \emph {et~al.}(2005)\citenamefont
  {Kollmann}, \citenamefont {Lovdok}, \citenamefont {Bartholome}, \citenamefont
  {Timmer},\ and\ \citenamefont {Sourjik}}]{kollmann_design_2005}%
  \BibitemOpen
  \bibfield  {author} {\bibinfo {author} {\bibfnamefont {M.}~\bibnamefont
  {Kollmann}}, \bibinfo {author} {\bibfnamefont {L.}~\bibnamefont {Lovdok}},
  \bibinfo {author} {\bibfnamefont {K.}~\bibnamefont {Bartholome}}, \bibinfo
  {author} {\bibfnamefont {J.}~\bibnamefont {Timmer}}, \ and\ \bibinfo {author}
  {\bibfnamefont {V.}~\bibnamefont {Sourjik}},\ }\href {\doibase
  10.1038/nature04228} {\bibfield  {journal} {\bibinfo  {journal} {Nature}\
  }\textbf {\bibinfo {volume} {438}},\ \bibinfo {pages} {504} (\bibinfo {year}
  {2005})}\BibitemShut {NoStop}%
\bibitem [{\citenamefont {Li}\ \emph {et~al.}(2004)\citenamefont {Li},
  \citenamefont {Long}, \citenamefont {Lu}, \citenamefont {Ouyang},\ and\
  \citenamefont {Tang}}]{li_yeast_2004}%
  \BibitemOpen
  \bibfield  {author} {\bibinfo {author} {\bibfnamefont {F.}~\bibnamefont
  {Li}}, \bibinfo {author} {\bibfnamefont {T.}~\bibnamefont {Long}}, \bibinfo
  {author} {\bibfnamefont {Y.}~\bibnamefont {Lu}}, \bibinfo {author}
  {\bibfnamefont {Q.}~\bibnamefont {Ouyang}}, \ and\ \bibinfo {author}
  {\bibfnamefont {C.}~\bibnamefont {Tang}},\ }\href {\doibase
  10.1073/pnas.0305937101} {\bibfield  {journal} {\bibinfo  {journal}
  {Proceedings of the National Academy of Sciences of the United States of
  America}\ }\textbf {\bibinfo {volume} {101}},\ \bibinfo {pages} {4781 }
  (\bibinfo {year} {2004})}\BibitemShut {NoStop}%
\bibitem [{\citenamefont {Albert}\ and\ \citenamefont
  {Othmer}(2003)}]{albert_topology_2003}%
  \BibitemOpen
  \bibfield  {author} {\bibinfo {author} {\bibfnamefont {R.}~\bibnamefont
  {Albert}}\ and\ \bibinfo {author} {\bibfnamefont {H.~G.}\ \bibnamefont
  {Othmer}},\ }\href {\doibase 10.1016/S0022-5193(03)00035-3} {\bibfield
  {journal} {\bibinfo  {journal} {Journal of Theoretical Biology}\ }\textbf
  {\bibinfo {volume} {223}},\ \bibinfo {pages} {1} (\bibinfo {year}
  {2003})}\BibitemShut {NoStop}%
\bibitem [{\citenamefont {Chaves}\ \emph {et~al.}(2005)\citenamefont {Chaves},
  \citenamefont {Albert},\ and\ \citenamefont
  {Sontag}}]{chaves_robustness_2005}%
  \BibitemOpen
  \bibfield  {author} {\bibinfo {author} {\bibfnamefont {M.}~\bibnamefont
  {Chaves}}, \bibinfo {author} {\bibfnamefont {R.}~\bibnamefont {Albert}}, \
  and\ \bibinfo {author} {\bibfnamefont {E.~D.}\ \bibnamefont {Sontag}},\
  }\href {\doibase 10.1016/j.jtbi.2005.01.023} {\bibfield  {journal} {\bibinfo
  {journal} {Journal of Theoretical Biology}\ }\textbf {\bibinfo {volume}
  {235}},\ \bibinfo {pages} {431} (\bibinfo {year} {2005})}\BibitemShut
  {NoStop}%
\bibitem [{\citenamefont {Maslov}\ and\ \citenamefont
  {Sneppen}(2005)}]{maslov_computational_2005}%
  \BibitemOpen
  \bibfield  {author} {\bibinfo {author} {\bibfnamefont {S.}~\bibnamefont
  {Maslov}}\ and\ \bibinfo {author} {\bibfnamefont {K.}~\bibnamefont
  {Sneppen}},\ }\href {http://www.iop.org/EJ/abstract/1478-3975/2/4/S03/}
  {\bibfield  {journal} {\bibinfo  {journal} {Physical Biology}\ }\textbf
  {\bibinfo {volume} {2}},\ \bibinfo {pages} {S94} (\bibinfo {year}
  {2005})}\BibitemShut {NoStop}%
\bibitem [{\citenamefont {Harris}\ \emph {et~al.}(2002)\citenamefont {Harris},
  \citenamefont {Sawhill}, \citenamefont {Wuensche},\ and\ \citenamefont
  {Kauffman}}]{harris_model_2002}%
  \BibitemOpen
  \bibfield  {author} {\bibinfo {author} {\bibfnamefont {S.~E.}\ \bibnamefont
  {Harris}}, \bibinfo {author} {\bibfnamefont {B.~K.}\ \bibnamefont {Sawhill}},
  \bibinfo {author} {\bibfnamefont {A.}~\bibnamefont {Wuensche}}, \ and\
  \bibinfo {author} {\bibfnamefont {S.}~\bibnamefont {Kauffman}},\ }\href
  {\doibase 10.1002/cplx.10022} {\bibfield  {journal} {\bibinfo  {journal}
  {Complexity}\ }\textbf {\bibinfo {volume} {7}},\ \bibinfo {pages} {23}
  (\bibinfo {year} {2002})}\BibitemShut {NoStop}%
\bibitem [{\citenamefont {Bornholdt}\ and\ \citenamefont
  {Sneppen}(1998)}]{bornholdt_neutral_1998}%
  \BibitemOpen
  \bibfield  {author} {\bibinfo {author} {\bibfnamefont {S.}~\bibnamefont
  {Bornholdt}}\ and\ \bibinfo {author} {\bibfnamefont {K.}~\bibnamefont
  {Sneppen}},\ }\href {\doibase 10.1103/PhysRevLett.81.236} {\bibfield
  {journal} {\bibinfo  {journal} {Physical Review Letters}\ }\textbf {\bibinfo
  {volume} {81}},\ \bibinfo {pages} {236} (\bibinfo {year} {1998})}\BibitemShut
  {NoStop}%
\bibitem [{\citenamefont {Bornholdt}\ and\ \citenamefont
  {Sneppen}(2000)}]{bornholdt_robustness_2000}%
  \BibitemOpen
  \bibfield  {author} {\bibinfo {author} {\bibfnamefont {S.}~\bibnamefont
  {Bornholdt}}\ and\ \bibinfo {author} {\bibfnamefont {K.}~\bibnamefont
  {Sneppen}},\ }\href {\doibase 10.1098/rspb.2000.1280} {\bibfield  {journal}
  {\bibinfo  {journal} {Proceedings of the Royal Society of London. Series B:
  Biological Sciences}\ }\textbf {\bibinfo {volume} {267}},\ \bibinfo {pages}
  {2281 } (\bibinfo {year} {2000})}\BibitemShut {NoStop}%
\bibitem [{\citenamefont {Stern}(1999)}]{stern_emergence_1999}%
  \BibitemOpen
  \bibfield  {author} {\bibinfo {author} {\bibfnamefont {M.~D.}\ \bibnamefont
  {Stern}},\ }\href {\doibase 10.1073/pnas.96.19.10746} {\bibfield  {journal}
  {\bibinfo  {journal} {Proceedings of the National Academy of Sciences}\
  }\textbf {\bibinfo {volume} {96}},\ \bibinfo {pages} {10746 } (\bibinfo
  {year} {1999})}\BibitemShut {NoStop}%
\bibitem [{\citenamefont {Bassler}\ \emph {et~al.}(2004)\citenamefont
  {Bassler}, \citenamefont {Lee},\ and\ \citenamefont
  {Lee}}]{bassler_evolution_2004}%
  \BibitemOpen
  \bibfield  {author} {\bibinfo {author} {\bibfnamefont {K.~E.}\ \bibnamefont
  {Bassler}}, \bibinfo {author} {\bibfnamefont {C.}~\bibnamefont {Lee}}, \ and\
  \bibinfo {author} {\bibfnamefont {Y.}~\bibnamefont {Lee}},\ }\href {\doibase
  10.1103/PhysRevLett.93.038101} {\bibfield  {journal} {\bibinfo  {journal}
  {Physical Review Letters}\ }\textbf {\bibinfo {volume} {93}},\ \bibinfo
  {pages} {038101} (\bibinfo {year} {2004})}\BibitemShut {NoStop}%
\bibitem [{\citenamefont {Aldana}\ \emph {et~al.}(2007)\citenamefont {Aldana},
  \citenamefont {Balleza}, \citenamefont {Kauffman},\ and\ \citenamefont
  {Resendiz}}]{aldana_robustness_2007}%
  \BibitemOpen
  \bibfield  {author} {\bibinfo {author} {\bibfnamefont {M.}~\bibnamefont
  {Aldana}}, \bibinfo {author} {\bibfnamefont {E.}~\bibnamefont {Balleza}},
  \bibinfo {author} {\bibfnamefont {S.}~\bibnamefont {Kauffman}}, \ and\
  \bibinfo {author} {\bibfnamefont {O.}~\bibnamefont {Resendiz}},\ }\href
  {\doibase 16/j.jtbi.2006.10.027} {\bibfield  {journal} {\bibinfo  {journal}
  {Journal of Theoretical Biology}\ }\textbf {\bibinfo {volume} {245}},\
  \bibinfo {pages} {433} (\bibinfo {year} {2007})}\BibitemShut {NoStop}%
\bibitem [{\citenamefont {Szejka}\ and\ \citenamefont
  {Drossel}(2007)}]{szejka_evolution_2007}%
  \BibitemOpen
  \bibfield  {author} {\bibinfo {author} {\bibnamefont {Szejka}}\ and\ \bibinfo
  {author} {\bibnamefont {Drossel}},\ }\href {\doibase
  10.1140/epjb/e2007-00135-2} {\bibfield  {journal} {\bibinfo  {journal} {Eur.
  Phys. J. B}\ }\textbf {\bibinfo {volume} {56}},\ \bibinfo {pages} {373}
  (\bibinfo {year} {2007})}\BibitemShut {NoStop}%
\bibitem [{\citenamefont {Mihaljev}\ and\ \citenamefont
  {Drossel}(2009)}]{mihaljev_evolution_2009}%
  \BibitemOpen
  \bibfield  {author} {\bibinfo {author} {\bibfnamefont {T.}~\bibnamefont
  {Mihaljev}}\ and\ \bibinfo {author} {\bibfnamefont {B.}~\bibnamefont
  {Drossel}},\ }\href {\doibase 10.1140/epjb/e2009-00032-8} {\bibfield
  {journal} {\bibinfo  {journal} {The European Physical Journal B}\ }\textbf
  {\bibinfo {volume} {67}},\ \bibinfo {pages} {259} (\bibinfo {year}
  {2009})}\BibitemShut {NoStop}%
\bibitem [{\citenamefont {Pomerance}\ \emph {et~al.}(2009)\citenamefont
  {Pomerance}, \citenamefont {Ott}, \citenamefont {Girvan},\ and\ \citenamefont
  {Losert}}]{pomerance_effect_2009}%
  \BibitemOpen
  \bibfield  {author} {\bibinfo {author} {\bibfnamefont {A.}~\bibnamefont
  {Pomerance}}, \bibinfo {author} {\bibfnamefont {E.}~\bibnamefont {Ott}},
  \bibinfo {author} {\bibfnamefont {M.}~\bibnamefont {Girvan}}, \ and\ \bibinfo
  {author} {\bibfnamefont {W.}~\bibnamefont {Losert}},\ }\href {\doibase
  10.1073/pnas.0900142106} {\bibfield  {journal} {\bibinfo  {journal}
  {Proceedings of the National Academy of Sciences}\ }\textbf {\bibinfo
  {volume} {106}},\ \bibinfo {pages} {8209} (\bibinfo {year}
  {2009})}\BibitemShut {NoStop}%
\bibitem [{Note1()}]{Note1}%
  \BibitemOpen
  \bibinfo {note} {Another realistic source of noise are perturbations in the
  update sequence of nodes, since gene regulation lacks a global synchronizing
  clock~\cite {klemm_topology_2005}. However, it can be shown that absolute
  robustness against this type of noise can be achieved in an independent
  manner, and with a very small effect to the global topological
  characteristics of the system~\cite {peixoto_boolean_2009}.}\BibitemShut
  {Stop}%
\bibitem [{\citenamefont {Peixoto}(2010)}]{peixoto_redundancy_2010}%
  \BibitemOpen
  \bibfield  {author} {\bibinfo {author} {\bibfnamefont {T.~P.}\ \bibnamefont
  {Peixoto}},\ }\href {\doibase 10.1103/PhysRevLett.104.048701} {\bibfield
  {journal} {\bibinfo  {journal} {Physical Review Letters}\ }\textbf {\bibinfo
  {volume} {104}},\ \bibinfo {pages} {048701} (\bibinfo {year}
  {2010})}\BibitemShut {NoStop}%
\bibitem [{\citenamefont {Peixoto}(2012)}]{peixoto_behavior_2012}%
  \BibitemOpen
  \bibfield  {author} {\bibinfo {author} {\bibfnamefont {T.~P.}\ \bibnamefont
  {Peixoto}},\ }\href {\doibase 10.1088/1742-5468/2012/01/P01006} {\bibfield
  {journal} {\bibinfo  {journal} {Journal of Statistical Mechanics: Theory and
  Experiment}\ }\textbf {\bibinfo {volume} {2012}},\ \bibinfo {pages} {P01006}
  (\bibinfo {year} {2012})}\BibitemShut {NoStop}%
\bibitem [{\citenamefont {Peixoto}\ and\ \citenamefont
  {Drossel}(2009{\natexlab{a}})}]{peixoto_noise_2009}%
  \BibitemOpen
  \bibfield  {author} {\bibinfo {author} {\bibfnamefont {T.~P.}\ \bibnamefont
  {Peixoto}}\ and\ \bibinfo {author} {\bibfnamefont {B.}~\bibnamefont
  {Drossel}},\ }\href {\doibase 10.1103/PhysRevE.79.036108} {\bibfield
  {journal} {\bibinfo  {journal} {Physical Review E}\ }\textbf {\bibinfo
  {volume} {79}},\ \bibinfo {pages} {036108} (\bibinfo {year}
  {2009}{\natexlab{a}})}\BibitemShut {NoStop}%
\bibitem [{\citenamefont {Nimwegen}(2006)}]{nimwegen_scaling_2006}%
  \BibitemOpen
  \bibfield  {author} {\bibinfo {author} {\bibfnamefont {E.}~\bibnamefont
  {Nimwegen}},\ }in\ \href
  {http://www.springerlink.com/content/x56481198w606w32/} {\emph {\bibinfo
  {booktitle} {Power Laws, {Scale-Free} Networks and Genome Biology}}}\
  (\bibinfo  {publisher} {Springer {US}},\ \bibinfo {address} {Boston, {MA}},\
  \bibinfo {year} {2006})\ pp.\ \bibinfo {pages} {236--253}\BibitemShut
  {NoStop}%
\bibitem [{Note2()}]{Note2}%
  \BibitemOpen
  \bibinfo {note} {The definition above will lead to a bias if $k_i$ is an even
  number, since if the sum happens to be exactly $k_i / 2$ the output will be
  $0$, arbitrarily. Alternative definitions could be used, which would remove
  the bias~\cite {szejka_phase_2008}. However, for the analysis presented here,
  this is not an issue since $k_i$ is always odd.}\BibitemShut {Stop}%
\bibitem [{\citenamefont {Evans}\ and\ \citenamefont
  {Schulman}(2003)}]{evans_maximum_2003}%
  \BibitemOpen
  \bibfield  {author} {\bibinfo {author} {\bibfnamefont {W.}~\bibnamefont
  {Evans}}\ and\ \bibinfo {author} {\bibfnamefont {L.}~\bibnamefont
  {Schulman}},\ }\href {\doibase 10.1109/TIT.2003.818405} {\bibfield  {journal}
  {\bibinfo  {journal} {{IEEE} Transactions on Information Theory}\ }\textbf
  {\bibinfo {volume} {49}},\ \bibinfo {pages} {3094} (\bibinfo {year}
  {2003})}\BibitemShut {NoStop}%
\bibitem [{\citenamefont {Huepe}\ and\ \citenamefont
  {{Aldana-González}}(2002)}]{huepe_dynamical_2002}%
  \BibitemOpen
  \bibfield  {author} {\bibinfo {author} {\bibfnamefont {C.}~\bibnamefont
  {Huepe}}\ and\ \bibinfo {author} {\bibfnamefont {M.}~\bibnamefont
  {{Aldana-González}}},\ }\href {\doibase 10.1023/A:1015777824097} {\bibfield
  {journal} {\bibinfo  {journal} {Journal of Statistical Physics}\ }\textbf
  {\bibinfo {volume} {108}},\ \bibinfo {pages} {527} (\bibinfo {year}
  {2002})}\BibitemShut {NoStop}%
\bibitem [{\citenamefont {Ispolatov}\ \emph {et~al.}(2005)\citenamefont
  {Ispolatov}, \citenamefont {Krapivsky},\ and\ \citenamefont
  {Yuryev}}]{ispolatov_duplication-divergence_2005}%
  \BibitemOpen
  \bibfield  {author} {\bibinfo {author} {\bibfnamefont {I.}~\bibnamefont
  {Ispolatov}}, \bibinfo {author} {\bibfnamefont {P.~L.}\ \bibnamefont
  {Krapivsky}}, \ and\ \bibinfo {author} {\bibfnamefont {A.}~\bibnamefont
  {Yuryev}},\ }\href {\doibase 10.1103/PhysRevE.71.061911} {\bibfield
  {journal} {\bibinfo  {journal} {Physical Review E}\ }\textbf {\bibinfo
  {volume} {71}},\ \bibinfo {pages} {061911} (\bibinfo {year}
  {2005})}\BibitemShut {NoStop}%
\bibitem [{\citenamefont {Enemark}\ and\ \citenamefont
  {Sneppen}(2007)}]{enemark_gene_2007}%
  \BibitemOpen
  \bibfield  {author} {\bibinfo {author} {\bibfnamefont {J.}~\bibnamefont
  {Enemark}}\ and\ \bibinfo {author} {\bibfnamefont {K.}~\bibnamefont
  {Sneppen}},\ }\href {http://adsabs.harvard.edu/abs/2007JSMTE..11....7E}
  {\bibfield  {journal} {\bibinfo  {journal} {Journal of Statistical Mechanics:
  Theory and Experiment}\ }\textbf {\bibinfo {volume} {11}},\ \bibinfo {pages}
  {7} (\bibinfo {year} {2007})}\BibitemShut {NoStop}%
\bibitem [{\citenamefont {Callen}(1985)}]{callen_thermodynamics_1985}%
  \BibitemOpen
  \bibfield  {author} {\bibinfo {author} {\bibfnamefont {H.~B.}\ \bibnamefont
  {Callen}},\ }\href@noop {} {\emph {\bibinfo {title} {Thermodynamics and an
  Introduction to Thermostatistics}}},\ \bibinfo {edition} {2nd}\ ed.\
  (\bibinfo  {publisher} {Wiley},\ \bibinfo {year} {1985})\BibitemShut
  {NoStop}%
\bibitem [{\citenamefont {Holland}\ \emph {et~al.}(1983)\citenamefont
  {Holland}, \citenamefont {Laskey},\ and\ \citenamefont
  {Leinhardt}}]{holland_stochastic_1983}%
  \BibitemOpen
  \bibfield  {author} {\bibinfo {author} {\bibfnamefont {P.~W.}\ \bibnamefont
  {Holland}}, \bibinfo {author} {\bibfnamefont {K.~B.}\ \bibnamefont {Laskey}},
  \ and\ \bibinfo {author} {\bibfnamefont {S.}~\bibnamefont {Leinhardt}},\
  }\href {\doibase 16/0378-8733(83)90021-7} {\bibfield  {journal} {\bibinfo
  {journal} {Social Networks}\ }\textbf {\bibinfo {volume} {5}},\ \bibinfo
  {pages} {109} (\bibinfo {year} {1983})}\BibitemShut {NoStop}%
\bibitem [{\citenamefont {Faust}\ and\ \citenamefont
  {Wasserman}(1992)}]{faust_blockmodels:_1992}%
  \BibitemOpen
  \bibfield  {author} {\bibinfo {author} {\bibfnamefont {K.}~\bibnamefont
  {Faust}}\ and\ \bibinfo {author} {\bibfnamefont {S.}~\bibnamefont
  {Wasserman}},\ }\href {\doibase 16/0378-8733(92)90013-W} {\bibfield
  {journal} {\bibinfo  {journal} {Social Networks}\ }\textbf {\bibinfo {volume}
  {14}},\ \bibinfo {pages} {5} (\bibinfo {year} {1992})}\BibitemShut {NoStop}%
\bibitem [{\citenamefont {Karrer}\ and\ \citenamefont
  {Newman}(2011)}]{karrer_stochastic_2011}%
  \BibitemOpen
  \bibfield  {author} {\bibinfo {author} {\bibfnamefont {B.}~\bibnamefont
  {Karrer}}\ and\ \bibinfo {author} {\bibfnamefont {M.~E.~J.}\ \bibnamefont
  {Newman}},\ }\href {\doibase 10.1103/PhysRevE.83.016107} {\bibfield
  {journal} {\bibinfo  {journal} {Physical Review E}\ }\textbf {\bibinfo
  {volume} {83}},\ \bibinfo {pages} {016107} (\bibinfo {year}
  {2011})}\BibitemShut {NoStop}%
\bibitem [{Note3()}]{Note3}%
  \BibitemOpen
  \bibinfo {note} {Blockmodels are essentially equivalent to the
  hidden-variable model~\cite {boguna_class_2003}, when the hidden variables
  are discrete, and their multiplicity is smaller than the number of
  nodes.}\BibitemShut {Stop}%
\bibitem [{\citenamefont {Newman}(2003{\natexlab{a}})}]{newman_structure_2003}%
  \BibitemOpen
  \bibfield  {author} {\bibinfo {author} {\bibfnamefont {M.~E.~J.}\
  \bibnamefont {Newman}},\ }\href@noop {} {\bibfield  {journal} {\bibinfo
  {journal} {{SIAM} Review}\ }\textbf {\bibinfo {volume} {45}},\ \bibinfo
  {pages} {167} (\bibinfo {year} {2003}{\natexlab{a}})}\BibitemShut {NoStop}%
\bibitem [{\citenamefont {Newman}(2003{\natexlab{b}})}]{newman_mixing_2003}%
  \BibitemOpen
  \bibfield  {author} {\bibinfo {author} {\bibfnamefont {M.~E.~J.}\
  \bibnamefont {Newman}},\ }\href
  {http://link.aps.org/abstract/PRE/v67/e026126} {\bibfield  {journal}
  {\bibinfo  {journal} {Phys. Rev. E}\ }\textbf {\bibinfo {volume} {67}},\
  \bibinfo {pages} {026126} (\bibinfo {year} {2003}{\natexlab{b}})}\BibitemShut
  {NoStop}%
\bibitem [{\citenamefont {Girvan}\ and\ \citenamefont
  {Newman}(2002)}]{girvan_community_2002}%
  \BibitemOpen
  \bibfield  {author} {\bibinfo {author} {\bibfnamefont {M.}~\bibnamefont
  {Girvan}}\ and\ \bibinfo {author} {\bibfnamefont {M.~E.~J.}\ \bibnamefont
  {Newman}},\ }\href {\doibase 10.1073/pnas.122653799} {\bibfield  {journal}
  {\bibinfo  {journal} {Proceedings of the National Academy of Sciences}\
  }\textbf {\bibinfo {volume} {99}},\ \bibinfo {pages} {7821 } (\bibinfo {year}
  {2002})}\BibitemShut {NoStop}%
\bibitem [{\citenamefont {Derrida}\ and\ \citenamefont
  {Pomeau}(1986)}]{derrida_random_1986}%
  \BibitemOpen
  \bibfield  {author} {\bibinfo {author} {\bibfnamefont {B.}~\bibnamefont
  {Derrida}}\ and\ \bibinfo {author} {\bibfnamefont {Y.}~\bibnamefont
  {Pomeau}},\ }\href@noop {} {\bibfield  {journal} {\bibinfo  {journal}
  {Europhys. Lett}\ }\textbf {\bibinfo {volume} {1}},\ \bibinfo {pages}
  {45–49} (\bibinfo {year} {1986})}\BibitemShut {NoStop}%
\bibitem [{\citenamefont {Bianconi}(2009)}]{bianconi_entropy_2009}%
  \BibitemOpen
  \bibfield  {author} {\bibinfo {author} {\bibfnamefont {G.}~\bibnamefont
  {Bianconi}},\ }\href {\doibase 10.1103/PhysRevE.79.036114} {\bibfield
  {journal} {\bibinfo  {journal} {Physical Review E}\ }\textbf {\bibinfo
  {volume} {79}},\ \bibinfo {pages} {036114} (\bibinfo {year}
  {2009})}\BibitemShut {NoStop}%
\bibitem [{\citenamefont {Peixoto}(2011)}]{peixoto_entropy_2011}%
  \BibitemOpen
  \bibfield  {author} {\bibinfo {author} {\bibfnamefont {T.~P.}\ \bibnamefont
  {Peixoto}},\ }\href {http://arxiv.org/abs/1112.6028} {\bibfield  {journal}
  {\bibinfo  {journal} {{arXiv:1112.6028}}\ } (\bibinfo {year}
  {2011})}\BibitemShut {NoStop}%
\bibitem [{Note4()}]{Note4}%
  \BibitemOpen
  \bibinfo {note} {This is not the only two-block structure which can generate
  arbitrary values of $b^*$. In~\cite {peixoto_behavior_2012} it is shown how a
  bipartite ``restoration'' structure also achieves this, albeit less
  efficiently.}\BibitemShut {Stop}%
\bibitem [{Note5()}]{Note5}%
  \BibitemOpen
  \bibinfo {note} {We have empirically verified this by minimizing the free
  energy with up to 10 blocks, and the solutions were always \protect \emph
  {identical} to that of the two-block case presented in the following section.
  The general character of the two-block topology was also verified by Monte
  Carlo simulations with up to $20$ blocks, as discussed later in the text (see
  also Fig.~\ref {fig:mc-multiblock}).}\BibitemShut {Stop}%
\bibitem [{\citenamefont {Byrd}\ \emph {et~al.}(1995)\citenamefont {Byrd},
  \citenamefont {Lu}, \citenamefont {Nocedal},\ and\ \citenamefont
  {Zhu}}]{byrd_limited_1995}%
  \BibitemOpen
  \bibfield  {author} {\bibinfo {author} {\bibfnamefont {R.~H.}\ \bibnamefont
  {Byrd}}, \bibinfo {author} {\bibfnamefont {P.}~\bibnamefont {Lu}}, \bibinfo
  {author} {\bibfnamefont {J.}~\bibnamefont {Nocedal}}, \ and\ \bibinfo
  {author} {\bibfnamefont {C.}~\bibnamefont {Zhu}},\ }\href {\doibase
  10.1137/0916069} {\bibfield  {journal} {\bibinfo  {journal} {{SIAM} Journal
  on Scientific Computing}\ }\textbf {\bibinfo {volume} {16}},\ \bibinfo
  {pages} {1190} (\bibinfo {year} {1995})}\BibitemShut {NoStop}%
\bibitem [{\citenamefont {Metropolis}\ \emph {et~al.}(1953)\citenamefont
  {Metropolis}, \citenamefont {Rosenbluth}, \citenamefont {Rosenbluth},
  \citenamefont {Teller},\ and\ \citenamefont
  {Teller}}]{metropolis_equation_1953}%
  \BibitemOpen
  \bibfield  {author} {\bibinfo {author} {\bibfnamefont {N.}~\bibnamefont
  {Metropolis}}, \bibinfo {author} {\bibfnamefont {A.~W.}\ \bibnamefont
  {Rosenbluth}}, \bibinfo {author} {\bibfnamefont {M.~N.}\ \bibnamefont
  {Rosenbluth}}, \bibinfo {author} {\bibfnamefont {A.~H.}\ \bibnamefont
  {Teller}}, \ and\ \bibinfo {author} {\bibfnamefont {E.}~\bibnamefont
  {Teller}},\ }\href {\doibase 10.1063/1.1699114} {\bibfield  {journal}
  {\bibinfo  {journal} {The Journal of Chemical Physics}\ }\textbf {\bibinfo
  {volume} {21}},\ \bibinfo {pages} {1087} (\bibinfo {year}
  {1953})}\BibitemShut {NoStop}%
\bibitem [{\citenamefont {Hastings}(1970)}]{hastings_monte_1970}%
  \BibitemOpen
  \bibfield  {author} {\bibinfo {author} {\bibfnamefont {W.~K.}\ \bibnamefont
  {Hastings}},\ }\href {\doibase 10.1093/biomet/57.1.97} {\bibfield  {journal}
  {\bibinfo  {journal} {Biometrika}\ }\textbf {\bibinfo {volume} {57}},\
  \bibinfo {pages} {97 } (\bibinfo {year} {1970})}\BibitemShut {NoStop}%
\bibitem [{Note6()}]{Note6}%
  \BibitemOpen
  \bibinfo {note} {One could argue that this may overlook the buildup of
  correlations in the network, since it assumes that the blocks are homogeneous
  and have an random in-degree distribution as given by Eq.~\ref {eq:pk}.
  However, we are \protect \emph {only} interested in networks which have this
  property, which, as discussed in the text, correspond to a partial
  maximization of entropy when the remaining constraints are in place, so this
  should not be an issue. To be sure, we have compared the value of $b^*$
  computed this way with the actual empirical value and found a very good
  agreement (see Fig.~\ref {fig:mc-trans}).}\BibitemShut {Stop}%
\bibitem [{\citenamefont {Abdulrehman}\ \emph {et~al.}(2010)\citenamefont
  {Abdulrehman}, \citenamefont {Monteiro}, \citenamefont {Teixeira},
  \citenamefont {Mira}, \citenamefont {Lourenco}, \citenamefont {dos Santos},
  \citenamefont {Cabrito}, \citenamefont {Francisco}, \citenamefont {Madeira},
  \citenamefont {Aires}, \citenamefont {Oliveira}, \citenamefont
  {{Sa-Correia}},\ and\ \citenamefont {Freitas}}]{abdulrehman_yeastract:_2010}%
  \BibitemOpen
  \bibfield  {author} {\bibinfo {author} {\bibfnamefont {D.}~\bibnamefont
  {Abdulrehman}}, \bibinfo {author} {\bibfnamefont {P.~T.}\ \bibnamefont
  {Monteiro}}, \bibinfo {author} {\bibfnamefont {M.~C.}\ \bibnamefont
  {Teixeira}}, \bibinfo {author} {\bibfnamefont {N.~P.}\ \bibnamefont {Mira}},
  \bibinfo {author} {\bibfnamefont {A.~B.}\ \bibnamefont {Lourenco}}, \bibinfo
  {author} {\bibfnamefont {S.~C.}\ \bibnamefont {dos Santos}}, \bibinfo
  {author} {\bibfnamefont {T.~R.}\ \bibnamefont {Cabrito}}, \bibinfo {author}
  {\bibfnamefont {A.~P.}\ \bibnamefont {Francisco}}, \bibinfo {author}
  {\bibfnamefont {S.~C.}\ \bibnamefont {Madeira}}, \bibinfo {author}
  {\bibfnamefont {R.~S.}\ \bibnamefont {Aires}}, \bibinfo {author}
  {\bibfnamefont {A.~L.}\ \bibnamefont {Oliveira}}, \bibinfo {author}
  {\bibfnamefont {I.}~\bibnamefont {{Sa-Correia}}}, \ and\ \bibinfo {author}
  {\bibfnamefont {A.~T.}\ \bibnamefont {Freitas}},\ }\href {\doibase
  10.1093/nar/gkq964} {\bibfield  {journal} {\bibinfo  {journal} {Nucleic Acids
  Research}\ }\textbf {\bibinfo {volume} {39}},\ \bibinfo {pages} {D136}
  (\bibinfo {year} {2010})}\BibitemShut {NoStop}%
\bibitem [{\citenamefont {{Gama-Castro}}\ \emph {et~al.}(2010)\citenamefont
  {{Gama-Castro}}, \citenamefont {Salgado}, \citenamefont {{Peralta-Gil}},
  \citenamefont {{Santos-Zavaleta}}, \citenamefont {{Muñiz-Rascado}},
  \citenamefont {{Solano-Lira}}, \citenamefont {{Jimenez-Jacinto}},
  \citenamefont {Weiss}, \citenamefont {{García-Sotelo}}, \citenamefont
  {{López-Fuentes}}, \citenamefont {{Porrón-Sotelo}}, \citenamefont
  {{Alquicira-Hernández}}, \citenamefont {{Medina-Rivera}}, \citenamefont
  {{Martínez-Flores}}, \citenamefont {{Alquicira-Hernández}}, \citenamefont
  {{Martínez-Adame}}, \citenamefont {{Bonavides-Martínez}}, \citenamefont
  {{Miranda-Ríos}}, \citenamefont {Huerta}, \citenamefont {{Mendoza-Vargas}},
  \citenamefont {{Collado-Torres}}, \citenamefont {Taboada}, \citenamefont
  {{Vega-Alvarado}}, \citenamefont {Olvera}, \citenamefont {Olvera},
  \citenamefont {Grande}, \citenamefont {Morett},\ and\ \citenamefont
  {{Collado-Vides}}}]{gama-castro_regulondb_2010}%
  \BibitemOpen
  \bibfield  {author} {\bibinfo {author} {\bibfnamefont {S.}~\bibnamefont
  {{Gama-Castro}}}, \bibinfo {author} {\bibfnamefont {H.}~\bibnamefont
  {Salgado}}, \bibinfo {author} {\bibfnamefont {M.}~\bibnamefont
  {{Peralta-Gil}}}, \bibinfo {author} {\bibfnamefont {A.}~\bibnamefont
  {{Santos-Zavaleta}}}, \bibinfo {author} {\bibfnamefont {L.}~\bibnamefont
  {{Muñiz-Rascado}}}, \bibinfo {author} {\bibfnamefont {H.}~\bibnamefont
  {{Solano-Lira}}}, \bibinfo {author} {\bibfnamefont {V.}~\bibnamefont
  {{Jimenez-Jacinto}}}, \bibinfo {author} {\bibfnamefont {V.}~\bibnamefont
  {Weiss}}, \bibinfo {author} {\bibfnamefont {J.~S.}\ \bibnamefont
  {{García-Sotelo}}}, \bibinfo {author} {\bibfnamefont {A.}~\bibnamefont
  {{López-Fuentes}}}, \bibinfo {author} {\bibfnamefont {L.}~\bibnamefont
  {{Porrón-Sotelo}}}, \bibinfo {author} {\bibfnamefont {S.}~\bibnamefont
  {{Alquicira-Hernández}}}, \bibinfo {author} {\bibfnamefont {A.}~\bibnamefont
  {{Medina-Rivera}}}, \bibinfo {author} {\bibfnamefont {I.}~\bibnamefont
  {{Martínez-Flores}}}, \bibinfo {author} {\bibfnamefont {K.}~\bibnamefont
  {{Alquicira-Hernández}}}, \bibinfo {author} {\bibfnamefont {R.}~\bibnamefont
  {{Martínez-Adame}}}, \bibinfo {author} {\bibfnamefont {C.}~\bibnamefont
  {{Bonavides-Martínez}}}, \bibinfo {author} {\bibfnamefont {J.}~\bibnamefont
  {{Miranda-Ríos}}}, \bibinfo {author} {\bibfnamefont {A.~M.}\ \bibnamefont
  {Huerta}}, \bibinfo {author} {\bibfnamefont {A.}~\bibnamefont
  {{Mendoza-Vargas}}}, \bibinfo {author} {\bibfnamefont {L.}~\bibnamefont
  {{Collado-Torres}}}, \bibinfo {author} {\bibfnamefont {B.}~\bibnamefont
  {Taboada}}, \bibinfo {author} {\bibfnamefont {L.}~\bibnamefont
  {{Vega-Alvarado}}}, \bibinfo {author} {\bibfnamefont {M.}~\bibnamefont
  {Olvera}}, \bibinfo {author} {\bibfnamefont {L.}~\bibnamefont {Olvera}},
  \bibinfo {author} {\bibfnamefont {R.}~\bibnamefont {Grande}}, \bibinfo
  {author} {\bibfnamefont {E.}~\bibnamefont {Morett}}, \ and\ \bibinfo {author}
  {\bibfnamefont {J.}~\bibnamefont {{Collado-Vides}}},\ }\href {\doibase
  10.1093/nar/gkq1110} {\bibfield  {journal} {\bibinfo  {journal} {Nucleic
  Acids Research}\ } (\bibinfo {year} {2010}),\
  10.1093/nar/gkq1110}\BibitemShut {NoStop}%
\bibitem [{\citenamefont {Balaji}\ \emph
  {et~al.}(2006{\natexlab{a}})\citenamefont {Balaji}, \citenamefont {Babu},
  \citenamefont {Iyer}, \citenamefont {Luscombe},\ and\ \citenamefont
  {Aravind}}]{balaji_comprehensive_2006}%
  \BibitemOpen
  \bibfield  {author} {\bibinfo {author} {\bibfnamefont {S.}~\bibnamefont
  {Balaji}}, \bibinfo {author} {\bibfnamefont {M.~M.}\ \bibnamefont {Babu}},
  \bibinfo {author} {\bibfnamefont {L.~M.}\ \bibnamefont {Iyer}}, \bibinfo
  {author} {\bibfnamefont {N.~M.}\ \bibnamefont {Luscombe}}, \ and\ \bibinfo
  {author} {\bibfnamefont {L.}~\bibnamefont {Aravind}},\ }\href {\doibase
  16/j.jmb.2006.04.029} {\bibfield  {journal} {\bibinfo  {journal} {Journal of
  Molecular Biology}\ }\textbf {\bibinfo {volume} {360}},\ \bibinfo {pages}
  {213} (\bibinfo {year} {2006}{\natexlab{a}})}\BibitemShut {NoStop}%
\bibitem [{\citenamefont {Cosentino~Lagomarsino}\ \emph
  {et~al.}(2007)\citenamefont {Cosentino~Lagomarsino}, \citenamefont {Jona},
  \citenamefont {Bassetti},\ and\ \citenamefont
  {Isambert}}]{cosentino_lagomarsino_hierarchy_2007}%
  \BibitemOpen
  \bibfield  {author} {\bibinfo {author} {\bibfnamefont {M.}~\bibnamefont
  {Cosentino~Lagomarsino}}, \bibinfo {author} {\bibfnamefont {P.}~\bibnamefont
  {Jona}}, \bibinfo {author} {\bibfnamefont {B.}~\bibnamefont {Bassetti}}, \
  and\ \bibinfo {author} {\bibfnamefont {H.}~\bibnamefont {Isambert}},\ }\href
  {\doibase 10.1073/pnas.0609023104} {\bibfield  {journal} {\bibinfo  {journal}
  {Proceedings of the National Academy of Sciences}\ }\textbf {\bibinfo
  {volume} {104}},\ \bibinfo {pages} {5516 } (\bibinfo {year}
  {2007})}\BibitemShut {NoStop}%
\bibitem [{\citenamefont {Fraser}\ \emph {et~al.}(2004)\citenamefont {Fraser},
  \citenamefont {Hirsh}, \citenamefont {Giaever}, \citenamefont {Kumm},\ and\
  \citenamefont {Eisen}}]{fraser_noise_2004}%
  \BibitemOpen
  \bibfield  {author} {\bibinfo {author} {\bibfnamefont {H.~B.}\ \bibnamefont
  {Fraser}}, \bibinfo {author} {\bibfnamefont {A.~E.}\ \bibnamefont {Hirsh}},
  \bibinfo {author} {\bibfnamefont {G.}~\bibnamefont {Giaever}}, \bibinfo
  {author} {\bibfnamefont {J.}~\bibnamefont {Kumm}}, \ and\ \bibinfo {author}
  {\bibfnamefont {M.~B.}\ \bibnamefont {Eisen}},\ }\href {\doibase
  10.1371/journal.pbio.0020137} {\bibfield  {journal} {\bibinfo  {journal}
  {{PLoS} Biol}\ }\textbf {\bibinfo {volume} {2}},\ \bibinfo {pages} {e137}
  (\bibinfo {year} {2004})}\BibitemShut {NoStop}%
\bibitem [{\citenamefont {Jothi}\ \emph {et~al.}(2009)\citenamefont {Jothi},
  \citenamefont {Balaji}, \citenamefont {Wuster}, \citenamefont {Grochow},
  \citenamefont {Gsponer}, \citenamefont {Przytycka}, \citenamefont {Aravind},\
  and\ \citenamefont {Babu}}]{jothi_genomic_2009}%
  \BibitemOpen
  \bibfield  {author} {\bibinfo {author} {\bibfnamefont {R.}~\bibnamefont
  {Jothi}}, \bibinfo {author} {\bibfnamefont {S.}~\bibnamefont {Balaji}},
  \bibinfo {author} {\bibfnamefont {A.}~\bibnamefont {Wuster}}, \bibinfo
  {author} {\bibfnamefont {J.~A.}\ \bibnamefont {Grochow}}, \bibinfo {author}
  {\bibfnamefont {J.}~\bibnamefont {Gsponer}}, \bibinfo {author} {\bibfnamefont
  {T.~M.}\ \bibnamefont {Przytycka}}, \bibinfo {author} {\bibfnamefont
  {L.}~\bibnamefont {Aravind}}, \ and\ \bibinfo {author} {\bibfnamefont
  {M.~M.}\ \bibnamefont {Babu}},\ }\href {\doibase 10.1038/msb.2009.52}
  {\bibfield  {journal} {\bibinfo  {journal} {Molecular Systems Biology}\
  }\textbf {\bibinfo {volume} {5}} (\bibinfo {year} {2009}),\
  10.1038/msb.2009.52}\BibitemShut {NoStop}%
\bibitem [{\citenamefont {Balaji}\ \emph
  {et~al.}(2006{\natexlab{b}})\citenamefont {Balaji}, \citenamefont {Iyer},
  \citenamefont {Aravind},\ and\ \citenamefont
  {Babu}}]{balaji_uncovering_2006}%
  \BibitemOpen
  \bibfield  {author} {\bibinfo {author} {\bibfnamefont {S.}~\bibnamefont
  {Balaji}}, \bibinfo {author} {\bibfnamefont {L.~M.}\ \bibnamefont {Iyer}},
  \bibinfo {author} {\bibfnamefont {L.}~\bibnamefont {Aravind}}, \ and\
  \bibinfo {author} {\bibfnamefont {M.~M.}\ \bibnamefont {Babu}},\ }\href
  {\doibase 16/j.jmb.2006.04.026} {\bibfield  {journal} {\bibinfo  {journal}
  {Journal of Molecular Biology}\ }\textbf {\bibinfo {volume} {360}},\ \bibinfo
  {pages} {204} (\bibinfo {year} {2006}{\natexlab{b}})}\BibitemShut {NoStop}%
\bibitem [{\citenamefont {Balazsi}\ \emph {et~al.}(2008)\citenamefont
  {Balazsi}, \citenamefont {Heath}, \citenamefont {Shi},\ and\ \citenamefont
  {Gennaro}}]{balazsi_temporal_2008}%
  \BibitemOpen
  \bibfield  {author} {\bibinfo {author} {\bibfnamefont {G.}~\bibnamefont
  {Balazsi}}, \bibinfo {author} {\bibfnamefont {A.~P.}\ \bibnamefont {Heath}},
  \bibinfo {author} {\bibfnamefont {L.}~\bibnamefont {Shi}}, \ and\ \bibinfo
  {author} {\bibfnamefont {M.~L.}\ \bibnamefont {Gennaro}},\ }\href {\doibase
  10.1038/msb.2008.63} {\bibfield  {journal} {\bibinfo  {journal} {Mol Syst
  Biol}\ }\textbf {\bibinfo {volume} {4}} (\bibinfo {year} {2008}),\
  10.1038/msb.2008.63}\BibitemShut {NoStop}%
\bibitem [{\citenamefont {Ma'ayan}\ \emph {et~al.}(2005)\citenamefont
  {Ma'ayan}, \citenamefont {Jenkins}, \citenamefont {Neves}, \citenamefont
  {Hasseldine}, \citenamefont {Grace}, \citenamefont {{Dubin-Thaler}},
  \citenamefont {Eungdamrong}, \citenamefont {Weng}, \citenamefont {Ram},
  \citenamefont {Rice}, \citenamefont {Kershenbaum}, \citenamefont
  {Stolovitzky}, \citenamefont {Blitzer},\ and\ \citenamefont
  {Iyengar}}]{maayan_formation_2005}%
  \BibitemOpen
  \bibfield  {author} {\bibinfo {author} {\bibfnamefont {A.}~\bibnamefont
  {Ma'ayan}}, \bibinfo {author} {\bibfnamefont {S.~L.}\ \bibnamefont
  {Jenkins}}, \bibinfo {author} {\bibfnamefont {S.}~\bibnamefont {Neves}},
  \bibinfo {author} {\bibfnamefont {A.}~\bibnamefont {Hasseldine}}, \bibinfo
  {author} {\bibfnamefont {E.}~\bibnamefont {Grace}}, \bibinfo {author}
  {\bibfnamefont {B.}~\bibnamefont {{Dubin-Thaler}}}, \bibinfo {author}
  {\bibfnamefont {N.~J.}\ \bibnamefont {Eungdamrong}}, \bibinfo {author}
  {\bibfnamefont {G.}~\bibnamefont {Weng}}, \bibinfo {author} {\bibfnamefont
  {P.~T.}\ \bibnamefont {Ram}}, \bibinfo {author} {\bibfnamefont {J.~J.}\
  \bibnamefont {Rice}}, \bibinfo {author} {\bibfnamefont {A.}~\bibnamefont
  {Kershenbaum}}, \bibinfo {author} {\bibfnamefont {G.~A.}\ \bibnamefont
  {Stolovitzky}}, \bibinfo {author} {\bibfnamefont {R.~D.}\ \bibnamefont
  {Blitzer}}, \ and\ \bibinfo {author} {\bibfnamefont {R.}~\bibnamefont
  {Iyengar}},\ }\href {\doibase 10.1126/science.1108876} {\bibfield  {journal}
  {\bibinfo  {journal} {Science}\ }\textbf {\bibinfo {volume} {309}},\ \bibinfo
  {pages} {1078} (\bibinfo {year} {2005})}\BibitemShut {NoStop}%
\bibitem [{\citenamefont {Ma'ayan}(2009)}]{maayan_insights_2009}%
  \BibitemOpen
  \bibfield  {author} {\bibinfo {author} {\bibfnamefont {A.}~\bibnamefont
  {Ma'ayan}},\ }\href {\doibase 10.1074/jbc.R800056200} {\bibfield  {journal}
  {\bibinfo  {journal} {Journal of Biological Chemistry}\ }\textbf {\bibinfo
  {volume} {284}},\ \bibinfo {pages} {5451} (\bibinfo {year}
  {2009})}\BibitemShut {NoStop}%
\bibitem [{\citenamefont {Blake}\ \emph {et~al.}(2006)\citenamefont {Blake},
  \citenamefont {Balázsi}, \citenamefont {Kohanski}, \citenamefont {Isaacs},
  \citenamefont {Murphy}, \citenamefont {Kuang}, \citenamefont {Cantor},
  \citenamefont {Walt},\ and\ \citenamefont {Collins}}]{blake_phenotypic_2006}%
  \BibitemOpen
  \bibfield  {author} {\bibinfo {author} {\bibfnamefont {W.~J.}\ \bibnamefont
  {Blake}}, \bibinfo {author} {\bibfnamefont {G.}~\bibnamefont {Balázsi}},
  \bibinfo {author} {\bibfnamefont {M.~A.}\ \bibnamefont {Kohanski}}, \bibinfo
  {author} {\bibfnamefont {F.~J.}\ \bibnamefont {Isaacs}}, \bibinfo {author}
  {\bibfnamefont {K.~F.}\ \bibnamefont {Murphy}}, \bibinfo {author}
  {\bibfnamefont {Y.}~\bibnamefont {Kuang}}, \bibinfo {author} {\bibfnamefont
  {C.~R.}\ \bibnamefont {Cantor}}, \bibinfo {author} {\bibfnamefont {D.~R.}\
  \bibnamefont {Walt}}, \ and\ \bibinfo {author} {\bibfnamefont {J.~J.}\
  \bibnamefont {Collins}},\ }\href {\doibase 10.1016/j.molcel.2006.11.003}
  {\bibfield  {journal} {\bibinfo  {journal} {Molecular Cell}\ }\textbf
  {\bibinfo {volume} {24}},\ \bibinfo {pages} {853} (\bibinfo {year}
  {2006})}\BibitemShut {NoStop}%
\bibitem [{Note7()}]{Note7}%
  \BibitemOpen
  \bibinfo {note} {Other evolutionary, non-equilibrium pathways are also
  possible, see e.g.~\cite {perotti_emergent_2009}.}\BibitemShut {Stop}%
\bibitem [{\citenamefont {Klemm}\ and\ \citenamefont
  {Bornholdt}(2005)}]{klemm_topology_2005}%
  \BibitemOpen
  \bibfield  {author} {\bibinfo {author} {\bibfnamefont {K.}~\bibnamefont
  {Klemm}}\ and\ \bibinfo {author} {\bibfnamefont {S.}~\bibnamefont
  {Bornholdt}},\ }\href {\doibase 10.1073/pnas.0509132102} {\bibfield
  {journal} {\bibinfo  {journal} {Proceedings of the National Academy of
  Sciences of the United States of America}\ }\textbf {\bibinfo {volume}
  {102}},\ \bibinfo {pages} {18414} (\bibinfo {year} {2005})}\BibitemShut
  {NoStop}%
\bibitem [{\citenamefont {Peixoto}\ and\ \citenamefont
  {Drossel}(2009{\natexlab{b}})}]{peixoto_boolean_2009}%
  \BibitemOpen
  \bibfield  {author} {\bibinfo {author} {\bibfnamefont {T.~P.}\ \bibnamefont
  {Peixoto}}\ and\ \bibinfo {author} {\bibfnamefont {B.}~\bibnamefont
  {Drossel}},\ }\href {\doibase 10.1103/PhysRevE.80.056102} {\bibfield
  {journal} {\bibinfo  {journal} {Physical Review E}\ }\textbf {\bibinfo
  {volume} {80}},\ \bibinfo {pages} {056102} (\bibinfo {year}
  {2009}{\natexlab{b}})}\BibitemShut {NoStop}%
\bibitem [{\citenamefont {Szejka}\ \emph {et~al.}(2008)\citenamefont {Szejka},
  \citenamefont {Mihaljev},\ and\ \citenamefont {Drossel}}]{szejka_phase_2008}%
  \BibitemOpen
  \bibfield  {author} {\bibinfo {author} {\bibfnamefont {A.}~\bibnamefont
  {Szejka}}, \bibinfo {author} {\bibfnamefont {T.}~\bibnamefont {Mihaljev}}, \
  and\ \bibinfo {author} {\bibfnamefont {B.}~\bibnamefont {Drossel}},\ }\href
  {\doibase 10.1088/1367-2630/10/6/063009} {\bibfield  {journal} {\bibinfo
  {journal} {New Journal of Physics}\ }\textbf {\bibinfo {volume} {10}},\
  \bibinfo {pages} {063009} (\bibinfo {year} {2008})}\BibitemShut {NoStop}%
\bibitem [{\citenamefont {Boguñá}\ and\ \citenamefont
  {{Pastor-Satorras}}(2003)}]{boguna_class_2003}%
  \BibitemOpen
  \bibfield  {author} {\bibinfo {author} {\bibfnamefont {M.}~\bibnamefont
  {Boguñá}}\ and\ \bibinfo {author} {\bibfnamefont {R.}~\bibnamefont
  {{Pastor-Satorras}}},\ }\href {\doibase 10.1103/PhysRevE.68.036112}
  {\bibfield  {journal} {\bibinfo  {journal} {Physical Review E}\ }\textbf
  {\bibinfo {volume} {68}},\ \bibinfo {pages} {036112} (\bibinfo {year}
  {2003})}\BibitemShut {NoStop}%
\bibitem [{\citenamefont {Perotti}\ \emph {et~al.}(2009)\citenamefont
  {Perotti}, \citenamefont {Billoni}, \citenamefont {Tamarit}, \citenamefont
  {Chialvo},\ and\ \citenamefont {Cannas}}]{perotti_emergent_2009}%
  \BibitemOpen
  \bibfield  {author} {\bibinfo {author} {\bibfnamefont {J.~I.}\ \bibnamefont
  {Perotti}}, \bibinfo {author} {\bibfnamefont {O.~V.}\ \bibnamefont
  {Billoni}}, \bibinfo {author} {\bibfnamefont {F.~A.}\ \bibnamefont
  {Tamarit}}, \bibinfo {author} {\bibfnamefont {D.~R.}\ \bibnamefont
  {Chialvo}}, \ and\ \bibinfo {author} {\bibfnamefont {S.~A.}\ \bibnamefont
  {Cannas}},\ }\href {\doibase 10.1103/PhysRevLett.103.108701} {\bibfield
  {journal} {\bibinfo  {journal} {Physical Review Letters}\ }\textbf {\bibinfo
  {volume} {103}},\ \bibinfo {pages} {108701} (\bibinfo {year}
  {2009})}\BibitemShut {NoStop}%
\end{thebibliography}%
\end{document}